\documentclass[fleqn,usenatbib]{mnras}

\usepackage{newtxtext,newtxmath}

\usepackage[T1]{fontenc}

\DeclareRobustCommand{\VAN}[3]{#2}
\let\VANthebibliography\thebibliography
\def\thebibliography{\DeclareRobustCommand{\VAN}[3]{##3}\VANthebibliography}

\usepackage{graphicx}	

\usepackage{siunitx}
\usepackage{natbib}
\usepackage{subcaption}
\usepackage{amsmath}	

\usepackage{amssymb}	
\usepackage{ulem}
\usepackage[dvipsnames]{xcolor}
\usepackage{anyfontsize}

\newcommand{\kms}{\mbox{$\>{\rm km\, s^{-1}}$}}

\newcommand{\kpc}{\mbox{$\>{\rm kpc}$}} 
\newcommand{\pc}{\mbox{$\>{\rm pc}$}} 
\newcommand{\Gyr}{\mbox{$\>{\rm Gyr}$}}

\newcommand{\Msun}{\>{\rm M_{\odot}}}

\newcommand\degrees{^\circ}
\newcommand{\avg}[1]{\mbox{$\left<{#1}\right>$}}

\newcommand{\sig}[1]{\mbox{$\sigma_{#1}$}}

\newcommand{\feh}{\mbox{$\rm [Fe/H]$}}

\newcommand{\gaia}{{\it Gaia}}

\newcommand{\dex}{\mbox{$\>{\rm dex}$}}

\newcommand{\Rhat}{\hat{\textbf{\textit{R}}}}

\newcommand{\ihat}{\hat{\textbf{\textit{i}}}}

\newcommand{\Rgc}{R_\mathrm{GC}}

\newcommand{\xbold}{\mathrm{\mathbf{x}}}
\newcommand{\vbold}{\mathrm{\mathbf{v}}}

\newcommand{\vrhat}{\mathbf{\hat{v}}_r}
\newcommand{\vlhat}{\mathbf{\hat{v}}_l}

\newcommand{\meanvr}{\mbox{\avg{v_r}}}
\newcommand{\meanvl}{\mbox{\avg{v_l}}}
\newcommand{\meanvR}{\mbox{\avg{v_R}}}
\newcommand{\meanvphi}{\mbox{\avg{v_\phi}}}

\newcommand{\covij}{\mbox{$\sigma_{ij}^2$}}
\newcommand{\vari}{\mbox{$\sigma_{ii}^2$}}
\newcommand{\varj}{\mbox{$\sigma_{jj}^2$}}
\newcommand{\aniij}{\mbox{$\beta_{ij}$}}
\newcommand{\corrij}{\mbox{$\rho_{ij}$}}

\newcommand{\vertexabs}{\mbox{$l_{\rm v}$}}
\newcommand{\vertex}{\mbox{$\Tilde{l}_{\rm v}$}}
\newcommand{\vertexabsRphi}{\mbox{$l^{R\phi}_{\rm v}$}}

\newcommand{\ani}{\mbox{$\beta_{rl}$}}
\newcommand{\corr}{\mbox{$\rho_{rl}$}}
\newcommand{\anicyl}{\mbox{$\beta_{R\phi}$}}
\newcommand{\corrcyl}{\mbox{$\rho_{R\phi}$}}

\def\eg{{\it e.g.}}

\def\ie{{\it i.e.}}

\def\aj{AJ}
\def\apjl{ApJL}
\def\apjs{ApJS}
\def\aap{A\&A}
\def\nat{Nature}

\defcitealias{gough-kelly+2022}{GK22}

\title[Bulge velocity ellipsoids]{The chemical and spatial variations of the bulge's velocity ellipsoids}

\author[San Martin Fernandez et al.]{Luis M. San Martin Fernandez$^{1}$\thanks{E-mail: luismi98sanmartin@gmail.com}, Steven Gough-Kelly$^{1}$, Victor P. Debattista$^{1}$\thanks{E-mail: vpdebattista@uclan.ac.uk},
\newauthor Oscar A. Gonzalez$^{2}$, Ilin Lazar$^{3}$, Alvaro Rojas-Arriagada$^{4,5,6,7}$, Leandro {Beraldo e Silva}$^{8}$
\\
$^{1}$Jeremiah Horrocks Institute, University of Central Lancashire, Preston, PR1 2HE, UK\\
$^{2}$UK Astronomy Technology Centre, Royal Observatory, Blackford Hill, Edinburgh, EH9 3HJ, UK\\
$^{3}$Centre for Astrophysics Research, School of Physics, Engineering \& Computer Science, University of Hertfordshire, Hatfield, AL10 9AB, UK\\
$^{4}$Departamento de F\'isica, Universidad de Santiago de Chile, Av. Victor Jara 3659, Santiago, Chile\\
$^{5}$Millennium Institute of Astrophysics, Avenida Vicuña Mackenna 4860, Macul, Santiago 82-0436, Chile\\
$^6$N\'ucleo Milenio ERIS\\
$^7$Center for Interdisciplinary Research in Astrophysics and Space Exploration (CIRAS), Universidad de Santiago de Chile, Santiago, Chile\\
$^8$ Department of Astronomy \& Steward Observatory, University of Arizona, Tucson, AZ, 85721, USA\\
}

\date{Accepted XXX. Received YYY; in original form ZZZ}

\pubyear{2024}

\begin{document}
\label{firstpage}
\pagerange{\pageref{firstpage}--\pageref{lastpage}}
\maketitle

\begin{abstract}
We study the velocity ellipsoids in an $N$-body$+$SPH simulation of a
barred galaxy which forms a bar with a BP bulge. We focus on the 2D
kinematics, and quantify the velocity ellipses by the anisotropy,
\aniij, the correlation, \corrij, and the vertex deviation,
\vertexabs. We explore the variations in these quantities based on
stellar age within the bulge and compare these results with the Milky
Way's bulge using data from APOGEE DR16 and {\it Gaia} DR3. We first
explore the variation of the model's velocity ellipses in
galactocentric velocities, $v_R$ and $v_\phi$, for two bulge
populations, a (relatively) young one and an old one. The bar imprints
quadrupoles on the distribution of ellipse properties, which are
stronger in the young population, as expected from their stronger
bar. The quadrupoles are distorted if we use heliocentric velocities
$v_r$ and $v_l$.  We then project these kinematics along the line of
sight onto the $(l,b)$-plane. Along the minor axis \ani\ changes from
positive at low $|b|$ to negative at large $|b|$, crossing over at
lower $|b|$ in the young stars. Consequently the vertex deviation
peaks at lower $|b|$ in the young population, but reaches similar peak
values in the old. The \corr\ is much stronger in the young stars, and
traces the bar strength. The APOGEE stars split by the median
\feh\ follow the same trends.  Lastly we explore the velocity ellipses
across the entire bulge region in $(l,b)$ space, finding good
qualitative agreement between the model and observations.
\end{abstract}




\section{Introduction}
\label{sec:intro}

Vertex deviation measures the tilt of the velocity ellipse away from the coordinate axes. In an axisymmetric system, symmetry requires that the vertex deviation is $0\degrees$. When computed for bulge stars along the minor (rotation) axis, a non-zero vertex deviation of the $v_r$-$v_l$ velocity ellipse is therefore a useful probe of the deviation of the bulge from axisymmetry, and in particular of the bar's triaxiality, even with a relatively small sample of stars. The measurement by \citet{zhao1994} of a significant vertex deviation for a sample of 62 K~giants in Baade's Window provided the first stellar kinematic evidence that the bulge of the Milky Way (MW) is barred. \citet{zhao1994} found that the vertex deviations of the metal-rich and metal-poor stars were perpendicular to one another, with negative and positive values respectively.
\citet{Soto_2007} studied a larger sample of stars and also found a negative vertex deviation for the metal-rich stars, but an untilted velocity ellipse for the metal-poor ones \citep[see also][]{soto2012}. They pointed out that the positive vertex deviation found by \citet{zhao1994} for metal-poor stars was mainly due to their small sample size, and was not statistically significant. These results were confirmed by \citet{Babusiaux2010}, who found a metal-rich population with significant vertex deviation, and a metal-poor population consistent with a spheroid of vanishing vertex deviation. \citet{simion2021} also found a metallicity dependence of the vertex deviation in Baade's Window, although they did not detect a coherent metallicity dependence in other fields.

The difference in the vertex deviation of metal-rich and metal-poor stars has been interpreted as evidence of the co-existence of two different components in the bulge.
In this scenario, the bulge is comprised of both a secularly evolved box/peanut bulge driven by the thickening of the bar \citep[e.g.][]{Sellwood2020}, as well as an older, nearly-axisymmetric, accreted "classical" bulge in the metal-poor stars \citep[e.g.][]{vasquez2013}. 
Other evidence that has been invoked in favour of a significant accreted population in the bulge include velocity dispersions that vary with metallicity \citep{zhao1994, babusiaux2016}, the weak bar in RR~Lyrae \citep{dekany+2013}, the vertical metallicity gradient \citep{zoccali2008, Gonzalez2011, Johnson_2011, johnson_2013}, the absence of an X-shape in metal-poor stars \citep{ness2012, uttenthaler2012, rojas-arriagada2014}, the bimodal metallicity distribution \citep{ness2013a, williams+2016, schultheis+2017, rojas_arriagada_2020, johnson+2022} and the mostly old bulge stellar ages \citep{ortolani+95, Kuijken2002, zoccali+2003, ferreras+2003, sahu+2006, clarkson2008, clarkson+2011, brown+2010, valenti+2013, Calamida2014, surot+2019, sit_ness2020}.

However \citet{Debattista2017} showed that many of the trends with metallicity found in the bulge can be reproduced purely by the secular evolution of the bar. This occurs because the bar exerts a different influence on stellar populations of different initial kinematics, spatially separating them. Cooler (hence relatively young\footnote{In this context, a young population can be just $1-2\Gyr$ younger than an ${\sim} 11\Gyr$-old population. Youth is very much relative for bulge populations.} and metal-rich) stars form a strong bar, with a clear X-shaped distribution, while hotter (hence older and metal-poor) stars form a weaker bar, which is thicker at the centre, and has little or no X-shape. The vertical metallicity gradient then results from the different density distribution of the young and old stars. Since the populations are separated by the bar on the basis of their kinematics, \citet{Debattista2017} referred to this process as {\it kinematic fractionation}. In this scenario the bimodal metallicity distribution requires contribution from a thick disc population \citep{Bekki2011, dimatteo2015, dimatteo2019, fragkoudi2017a, fragkoudi2017b, fragkoudi2018, fragkoudi2020}, possibly via the accretion of star forming clumps \citep{debattista+23}, which is also an in-situ formation mechanism. Observations of high redshift galaxies also find evidence of an early formation of bulges via the gravitational instability of the gas and subsequent bursty star formation \citep[e.g.][]{tacchella2015, Tadaki_2017}.  \citet{shen2010} used a set of $N$-body simulations to show that the kinematics of the bulge can be matched with a bulge that included no more than $8\%$ of the total mass in the form of a slowly rotating classical bulge. \citet{Debattista2017} refined this estimate by considering the kinematics as a function of metallicity, finding that a classical bulge needed to be less massive than ${\sim} 2\%$ of the total stellar mass.

\citet{Debattista2019} studied the variation of the vertex deviation with age and metallicity in a cosmological simulation of a MW-like galaxy from the FIRE project \citep{hopkins+2015, hopkins+2018}. Rescaling the simulation, they found, for the equivalent of Baade's Window, a vertex deviation trend with metallicity comparable to that of \citet{Soto_2007} and \citet{Babusiaux2010}: a significant and constant vertex deviation for stars with $\feh\gtrsim-0.7$ and a vanishing one for metal-poor stars. 
They showed that the vertex deviation of old stars is low regardless of whether all stars are included or just those that formed in situ, suggesting that vertex deviation, as indeed many other kinematic quantities \citep{gough-kelly2022}, exhibits the effect of kinematic fractionation.

In this study, we analyse the kinematics of a high-resolution $N$-body$+$SPH simulation where stars form continuously out of gas, and compare the results with the \citet{rojas_arriagada_2020} sample of APOGEE-1 and APOGEE-2 stars \citep{Majewski_2017}, with proper motions from {\it Gaia} DR3 \citep{gaiaDR3}. Our model is the same as that used in \citet{Debattista2017}, which evolves in isolation purely secularly and exhibits kinematic fractionation. This high resolution simulation allows us to predict trends of the velocity ellipses across the bulge region. A qualitative agreement with the MW's trends would constitute evidence that our Galaxy's bulge formed secularly with stellar populations separated by kinematic fractionation, without the need to invoke a significant accreted bulge component. 

We first define the vertex deviation in
Section~\ref{sec:vd}. Section~\ref{sec:simulation} describes the
simulation from which the model we use is drawn. In
Section~\ref{sec:faceon_maps} we map the model's bulge galactocentric
and heliocentric kinematics, paying attention not just to the vertex
deviation but also to the in-plane anisotropy and
correlation. Section~\ref{sec:data} describes the observational data
we use to compare with the model. Then in
Section~\ref{sec:minor_axis_comparison} we explore the vertex
deviation along the bulge minor axis, comparing the heliocentric
kinematics of the model and of the observations. We explore the
kinematics both as a function of latitude, and as a function of age
(for the model) or metallicity (for the observations). In
Section~\ref{sec:lb_maps} we extend our study of the ellipses across
the entire bulge, for both the model and the observations. In
Section~\ref{sec:discussion} we discuss and summarise our results. In
the appendices, we rederive the vertex deviation and contrast two
differing definitions, briefly present results for two further
simulations, and discuss the validity of bootstrapping for our
statistics of interest.


\section{Vertex deviation}
\label{sec:vd}

The spread in stellar velocities at each position $\xbold$ in the Galaxy around the mean velocity is described by the velocity-dispersion tensor,
\begin{align}
    \sigma_{ij}^2(\xbold) &\equiv \frac{1}{\rho(\xbold)}\int (v_i-\langle v_i\rangle)(v_j-\langle v_j \rangle)f(\xbold,\vbold) \mathrm{d}^3\vbold \nonumber \\
    &=\langle v_i v_j \rangle - \langle v_i \rangle \langle v_j \rangle,
\label{eq:dispersion_tensor}
\end{align}
where $i,j\in{\{1,2,3\}}$ are 3D velocity components, $f(\xbold,\vbold)$ is the stellar distribution function at the phase space position $(\xbold,\vbold)$, and ${\rho(\xbold)=\int f(\xbold,\vbold)d^3 \vbold}$ is the density.

Since the velocity dispersion tensor is symmetric, it can always be diagonalised. The principal axes of this tensor define a 2D projection of the velocity ellipsoid, \ie\ a velocity ellipse. The deviation of the velocity ellipse from alignment with the coordinate axes is called the vertex deviation, $\vertexabs$, and is given by
\begin{equation}
    \vertexabs = \frac{1}{2}\arctan\frac{2\covij}{|\vari -\varj|},
\label{eq:standard_abs}
\end{equation}
which takes values $-45\degrees \leq l_v \leq 45\degrees$ (see Appendix \ref{vd_appendix}). In order to interpret vertex deviation values, it is useful to separately study each of the quantities that appear in Eqn.~\ref{eq:standard_abs}. Instead of the covariance ($\covij$) and variances ($\vari$, $\varj$) of the velocities, we consider their dimensionless equivalents. These are the correlation,
\begin{equation}
    \corrij = \frac{\covij}{\sigma_i \sigma_j},
    \label{eq:corr}
\end{equation}
which takes values $-1 \leq \rho_{ij} \leq 1$, and the in-plane anisotropy,
\begin{equation}
    \aniij = 1 - \frac{\varj}{\vari},
    \label{eq:anisotropy}
\end{equation}
which takes values $-\infty \leq \beta_{ij} \leq 1$. The anisotropy is negative when ${\vari < \varj}$. Then \vertexabs\ can be expressed purely in terms of \corrij\ and \aniij\ (see Appendix~\ref{appendix:anicorr}, Eqn.~\ref{eq:standard_abs_anicorr}).

Eqn.~\ref{eq:standard_abs} measures the degree by which the local direction of highest dispersion deviates from alignment with the coordinate axes with respect to which the velocities are being measured. The direction of highest dispersion in an axisymmetric bulge is everywhere cylindrically radial, \ie\ aligned along $\Rhat$ of galactocentric cylindrical coordinates. The presence of a bar leads to a deviation from such alignment, because the bar-supporting orbits stream along the bar. The vertex deviation can detect this deviation, making it a useful tool for studying the bar. Thus, for instance, \citet{simion2021} used the vertex deviation and the simulation of \citet{shen2010} to show that the Milky Way's bar is tilted by $29\degrees \pm 3\degrees$ with respect to the line joining the Sun and the Galactic Centre. 

Vertex deviation quantifies non-axisymmetry when stars are selected in a region where the chosen coordinate system captures the symmetry of an axisymmetric galaxy. Doing so ensures that a deviation of the direction of highest dispersion from the coordinate axes (as measured by vertex deviation) can be interpreted as a deviation from axisymmetry. The Galactocentric cylindrical coordinate system fulfills this everywhere across the Galaxy. However, the Galactocentric cylindrical unit vectors change direction rapidly across the bulge region due to proximity to the Galactic center. Given stars selected to build a velocity ellipse have to ``agree'' on the directions of the coordinate axes for the vertex deviation value to be meaningful, a vertex deviation analysis using the Galactocentric cylindrical coordinate system requires dense spatial coverage and high spatial resolution (necessitating observations with very precise heliocentric distances). The spherical Galactic coordinate system, on the contrary, has unit vectors whose directions are relatively stable across the bulge region but requires stars to be selected near $l=0\degrees$, where the unit vectors align most closely with those of the Galactocentric cylidrical coordinate system. Studies of the $v_r$-$v_l$ velocity ellipse in Baade's Window have been preferred not only because of its low extinction but also because it is close to $l=0\degrees$. Elsewhere the spherical Galactic unit vectors are a combination of Galactocentric radial and tangential velocities, whose directions vary with heliocentric distance, which complicates the interpretation of the vertex deviation values.

Eqn.~\ref{eq:standard_abs} shows that the vertex deviation becomes increasingly sensitive to any non-zero correlation as the anisotropy approaches zero. What happens in such cases is that, for an isotropic group of stars, given the lack of preferential direction towards either coordinate axis, any non-zero correlation by symmetry results in an ellipse that is elongated along the diagonal. Mathematically, $\vertexabs \to \operatorname{sign}(\corrij)\cdot45\degrees$ as $\aniij\to 0$ if $\corrij \neq 0$. The lack of dependence on the magnitude of $\corrij$ in such situations means that vertex deviation is a blunt probe of non-axisymmetry when the velocity distribution is isotropic. If the correlation is instead negligible, the velocity ellipse of the isotropic population approximates a circle, with poorly-defined semi-major axis direction. As a result, its vertex deviation is unstable and exhibits no statistical significance upon bootstrapping (see Eqn.~\ref{eq:bootstrap}). That is the special case of isotropic axisymmetry.

We will reserve the symbol $\vertexabs$ to refer to the vertex deviation of the velocity ellipse constructed using the heliocentric spherical Galactic velocities $v_r$ and $v_l$. The use of other coordinate systems will be reflected in the notation. For example, $\vertexabsRphi$ will be computed using the Galactocentric cylindrical velocities $v_R$ and $v_\phi$.


\section{Simulation}
\label{sec:simulation}

\begin{figure}
    \centering
    \includegraphics[width=\columnwidth]{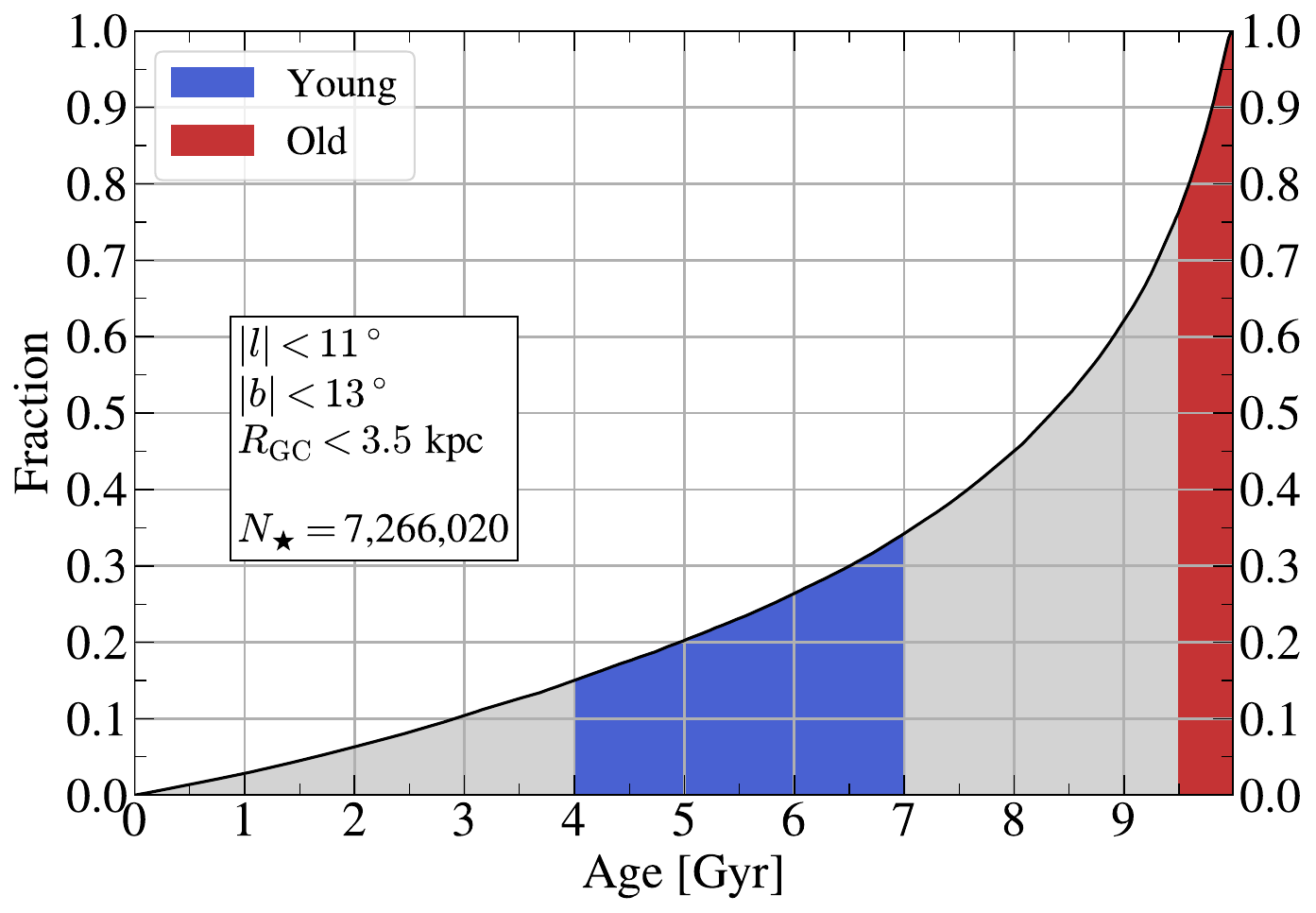}
    \caption{Cumulative distribution of the stellar ages in the model for the bulge region. The total number of bulge stars is ${N_\bigstar \sim7.27\times10^6}$. We take stars in the range $4$-$7\Gyr$ (${\sim}19\%$) and $9.5$-$10\Gyr$ (${\sim}24\%$) as representing our young and old bulge populations, respectively.}
    \label{fig:cumu_age}
\end{figure}

We study the dependence of the vertex deviation on stellar ages across the bulge using the same $N$-body$+$smooth particle hydrodynamics (SPH) simulation as was used by \citet{Debattista2017} to demonstrate kinematic fractionation. The simulation is evolved in isolation with {\sc gasoline} \citep{wadsley2004} for $10 \Gyr$, by which point it has formed ${\sim} 1.1\times 10^7$ stellar particles (hereafter stars). We use the $10\Gyr$ snapshot as our model. This same model was also used by \citet{gough-kelly2022} to compare the bulge proper motion rotation curves of old and young stars; our use of the model is very similar to that of \citet{gough-kelly2022}. Besides these works, the same model and the associated simulation has also been presented in several other studies \citep[\eg][]{Gardner2014, Cole2014, Ness2014, Gonzalez2016}, therefore we do not describe it at length here, but merely refer the reader to those earlier papers.

The simulation evolves via the cooling of gas off a hot corona in pressure equilibrium within a spherical dark matter halo. The dark matter halo has a virial radius of ${\sim} 200\kpc$, a virial mass of $9\times 10^{11}\Msun$, and a concentration $c=19$; the halo is represented by 5 million particles of unequal mass as described in \citet{Cole2014}. The gas follows the same radial profile but has only $11\%$ of the total mass. The gas is imparted with a solid body rotation to give it a spin $\lambda = 0.041$. Initially the gas corona is also comprised of 5 million particles (for an initial gas particle mass of $2.7\times 10^4\Msun$). As the gas cools, it settles into a disc and, where it reaches high density ($>100$ amu cm$^{-3}$), it ignites star formation with a probability of $0.1$ if the gas is part of a convergent flow and cooler than $1.5 \times 10^4$~K. Star formation then leads to supernova feedback using the blastwave prescription of \citet{Stinson2006}. 

The stars have an initial mass of $9.4 \times 10^3\Msun$ (up to $46\%$ of the mass of a star is lost via supernova feedback and AGB stellar winds), leading to a total stellar mass at $10\Gyr$ of $6.5 \times 10^{10}\Msun$. Supernova explosions couple $0.4 \times 10^{51}$ ergs per supernova \citep{governato2010} to the gas, and metals pollute the interstellar medium using the yields of \citet{raiteri+1996} for SN~II, 
\citet{thielemann+1986} for SN~Ia, and \citet{weidemann+1987} for stellar winds. However, the simulation does not include the diffusion of metals between the gas particles, which results in an excess of low-metallicity stars forming at all ages, and a weakened relation between metallicity and age. As a result, following previous works \citep{Debattista2017, gough-kelly2022}, we define stellar populations on the basis of age rather than metallicity. As the age is more fundamental anyway (albeit harder to measure observationally), this is an acceptable trade-off.

\subsection{The model bar}
\label{subsec:model-bar}

A bar forms in the simulation during the time interval $2-4\Gyr$ and grows secularly thereafter \citep{Cole2014}. At 2\Gyr\ around $55\%$ of the total stellar mass in the galaxy has already formed, which is not unusual for galaxies of Sb-type \citep{tacchella2015}. The result is a bulge with mostly old stars, as shown in the cumulative distribution of ages in Fig.~\ref{fig:cumu_age}, consistent with observations of the MW bulge \citep[\eg][]{Kuijken2002,clarkson2008,valenti2013,renzini2018, surot+2019}.

\begin{figure*}
    \begin{subfigure}{\textwidth}
        \centering
        \includegraphics[width=0.96\textwidth]{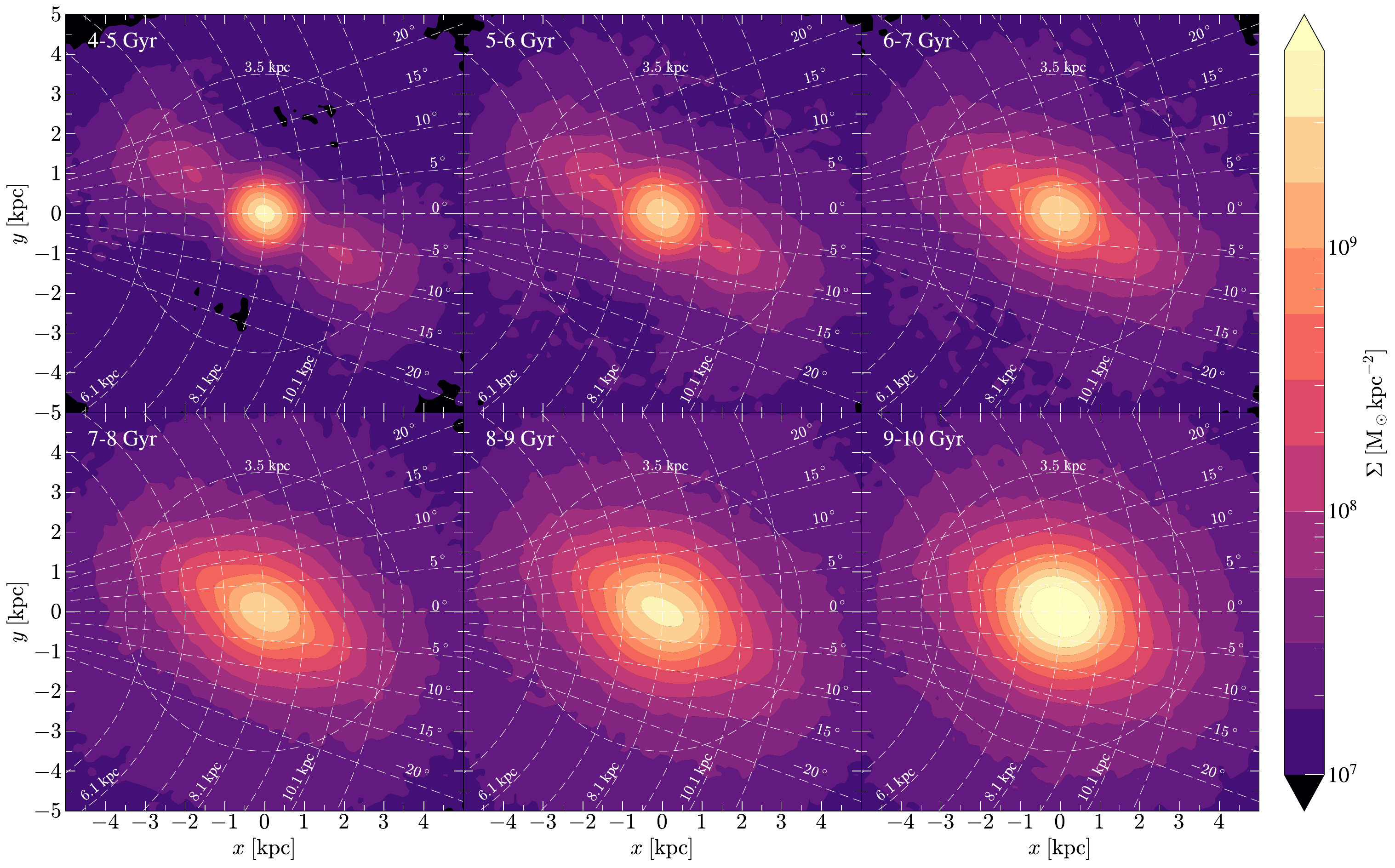}
        \caption{$|z|<3\kpc$}
        \label{subfig:top_age_windows_min0}
    \end{subfigure}
    \begin{subfigure}{\textwidth}
        \centering
        \includegraphics[width=0.96\textwidth]{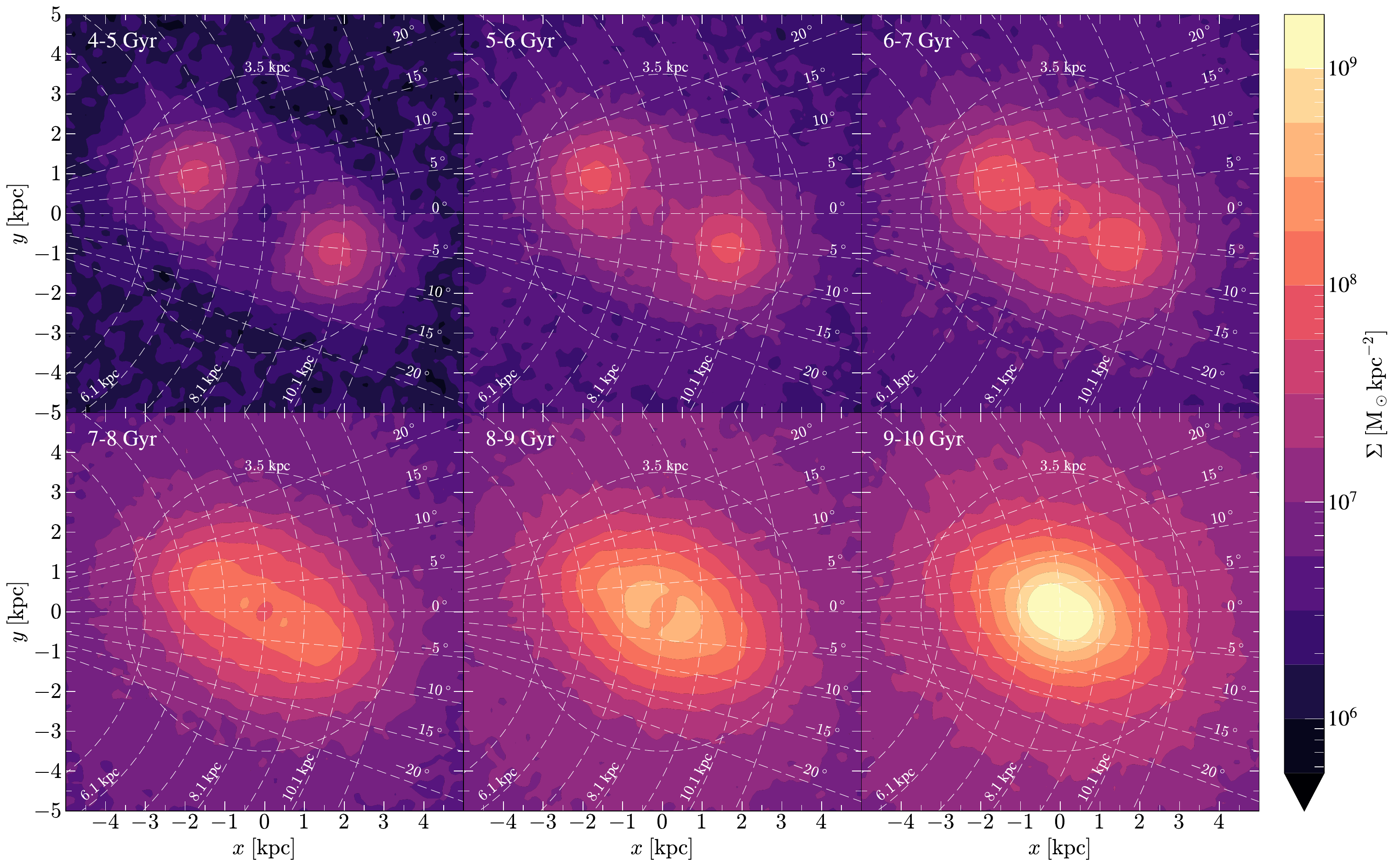}
        \caption{$0.5<|z|<3\kpc$}
        \label{subfig:top_age_windows_min0.5}
    \end{subfigure}
    \caption{Face-on view of stars of different ages, indicated at top left of each panel, in the bulge of the model, with (a) $|z|< 3\kpc$ and (b) $0.5<|z|<3 \kpc$. The Sun is located at $(x,y) = (-8.1,0)\kpc$. White dashed lines represent longitudes from $-20\degrees$ to $20\degrees$ in steps of $5\degrees$ (straight lines) and distances from the Sun from $5.1$ to $11.1\kpc$ in steps of $1\kpc$ (curves). The $\Rgc=3.5\kpc$ circle is also shown.}
    \label{fig:top_age_windows}
\end{figure*}

\begin{figure*}
    \centering
    \includegraphics[width=\textwidth]{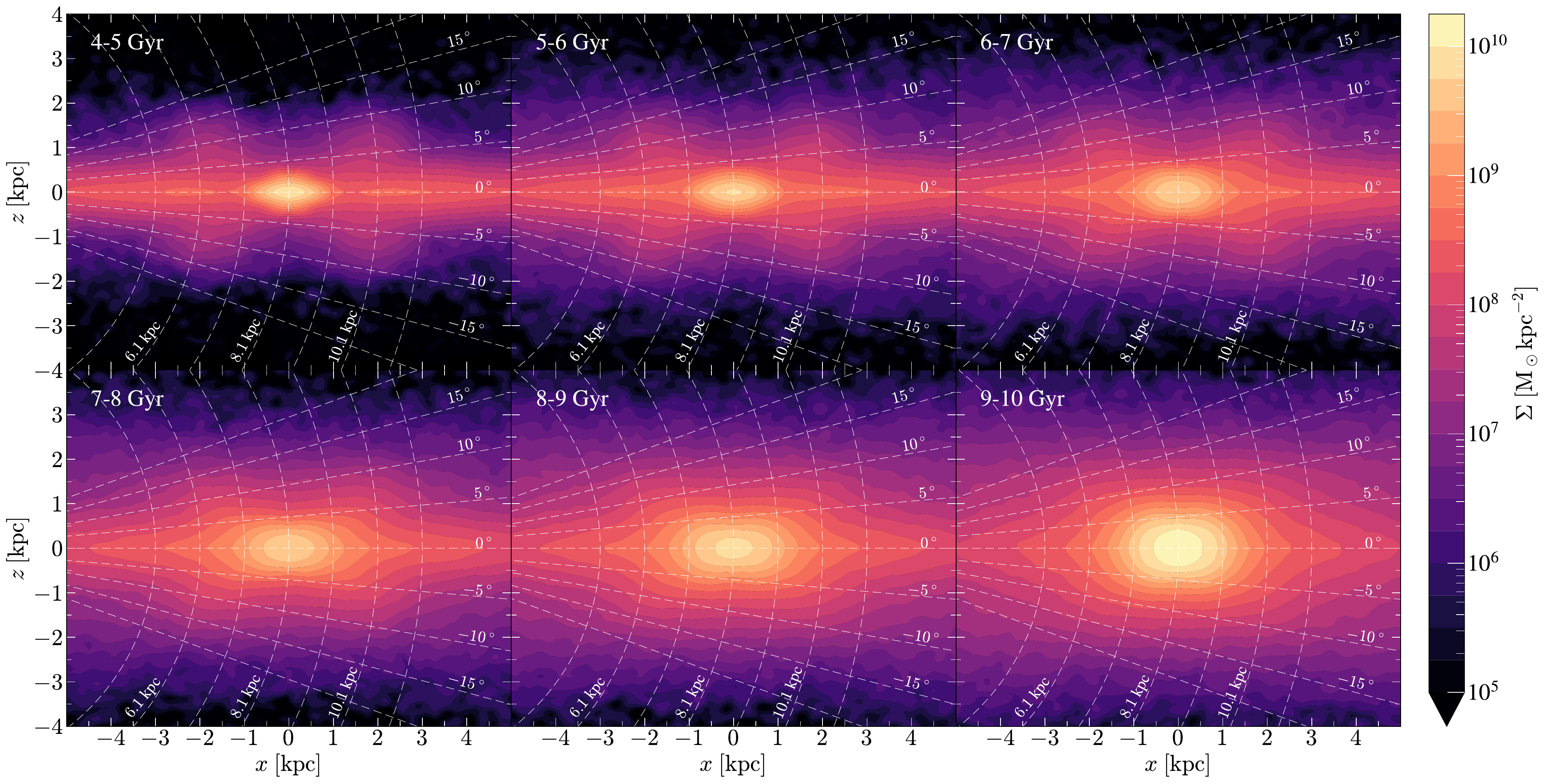}
    \caption{Edge-on view of stars of different ages, as indicated at top left in each panel, in the bulge of the model, with $|y|< 5 \kpc$. The Sun is located at $(x,y) = (-8.1,0)\kpc$. White dashed lines represent latitudes from $-20\degrees$ to $20\degrees$ in steps of $5\degrees$ (straight lines) and distances from the Sun from $5.1$ to $11.1\kpc$ in steps of $1\kpc$ (curves).}
    \label{fig:side_age_windows}
\end{figure*}

The model bar has a semi-major axis of ${\sim}2.9\kpc$ \citep{gough-kelly2022}, whereas in the MW \citet{Wegg2015} measured a bar semi-major axis of ${\sim} 5\kpc$. We therefore apply a scaling factor of 1.7 to the spatial coordinates of the model. We also apply a kinematic scaling of 0.48 to the model's velocities to scale it to the MW, as in previous studies \citep{Debattista2017, gough-kelly2022}; however this has no effect on the anisotropy, correlation, and vertex deviation since these are based on dimensionless ratios.

Starting with the bar aligned with the $x$-axis, we rotate it clockwise by $27\degrees$ \citep{Wegg2013}, with the receding near end at positive longitudes. We adopt a right-handed galactocentric rectangular coordinate system, and place the observer on the negative $x$-axis at $R_0=8.1\kpc$ from the galactic centre \citep{gravity2018,Qin2018}.
In order to match our observational sample (presented in Section \ref{sec:data}), we define the bulge as the region enclosed by ${|l|<11\degrees}$, ${|b|<13\degrees}$, and ${\Rgc<3.5\kpc}$. This results in a total of $7{,}266{,}020$ bulge stars.

The face-on view of the stellar populations of different ages in the bulge is shown in Fig.~\ref{fig:top_age_windows}, with (a) $|z|<3\kpc$ and (b) $0.5<|z|/\kpc<3$. We show stars of ages from $4$ to $10\Gyr$ in intervals of $1\Gyr$. The side-on view of the same stellar populations, with $|y|<5\kpc$, is shown in Fig.~\ref{fig:side_age_windows}.

Fig.~\ref{fig:top_age_windows} shows the elongated shape of the distribution of stars older than 4\Gyr, which diminishes progressively as we approach 10\Gyr. Similarly, in Fig.~\ref{fig:side_age_windows}, the distribution of younger stars exhibits a prominent X-shape, while that of old stars is boxy. At $|b|\gtrsim5\degrees$ the arms of the X-shape cause a double peak in the distance distribution \citep{Debattista2017, gough-kelly2022}, consistent with the red clump bimodality found in the MW \citep{McWilliam2010, Saito2011}.

\begin{figure*}
    \begin{subfigure}{\textwidth}
        \centering
        \includegraphics[width=\textwidth]{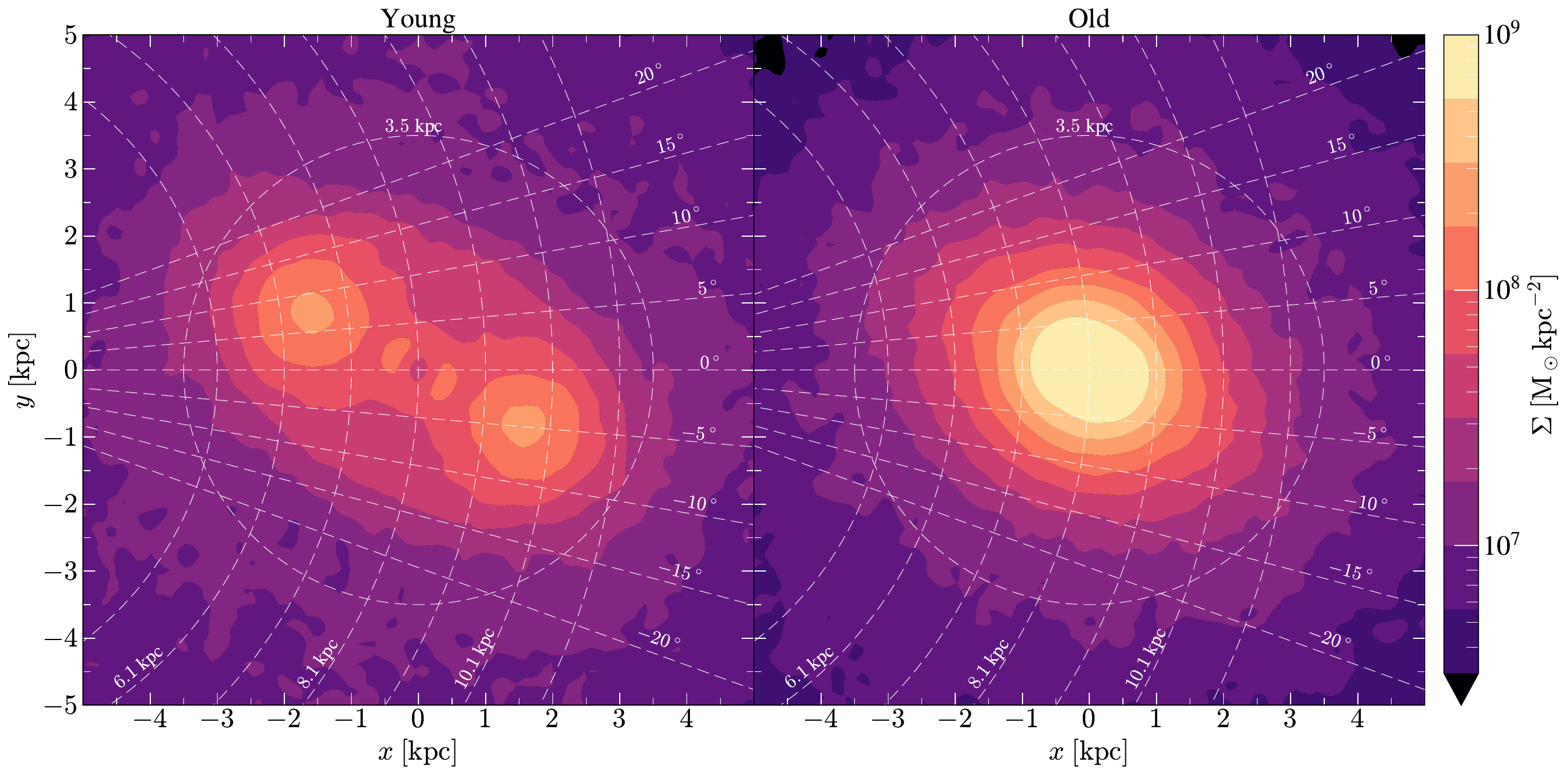}
        \caption{Face-on}        \label{subfig:top_density_youngold}
    \end{subfigure}
    \begin{subfigure}{\textwidth}
        \centering
        \includegraphics[width=\textwidth]{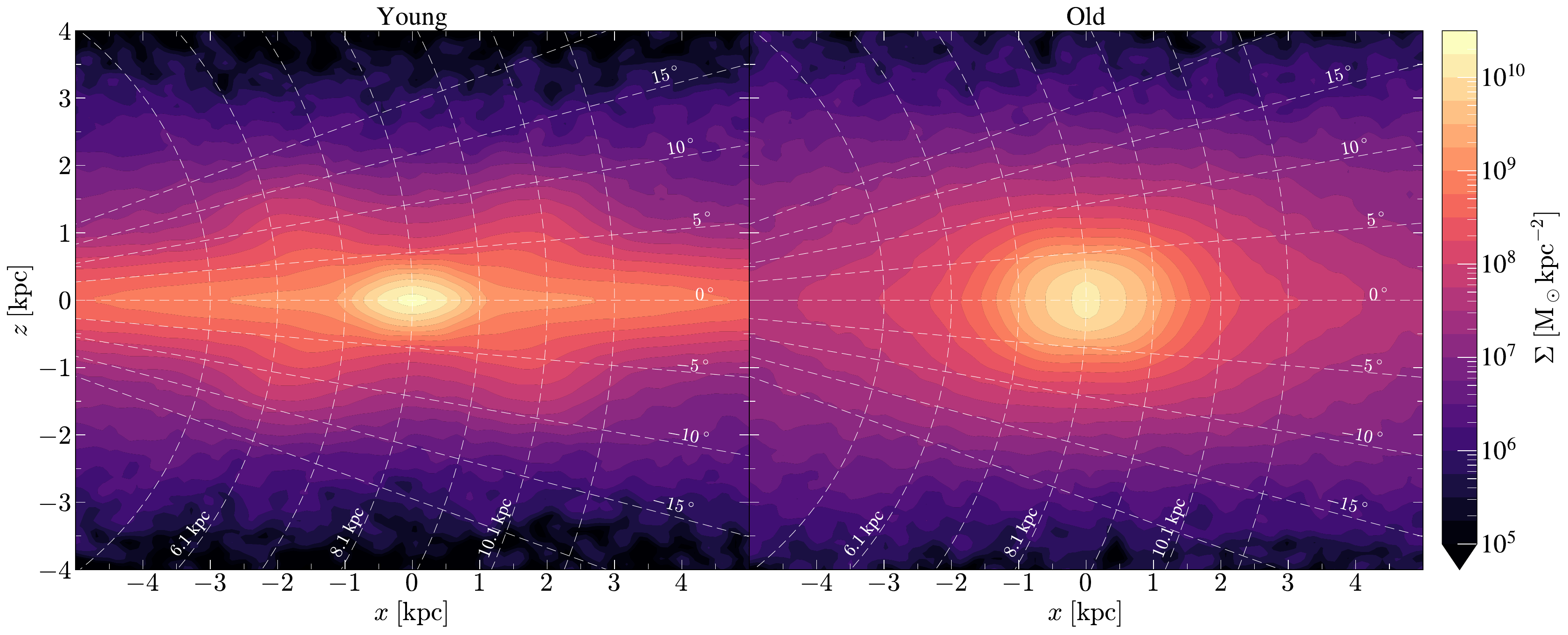}
        \caption{Side-on}        \label{subfig:side_density_youngold}
    \end{subfigure}
    \caption{(a) Face-on and (b) side-on views of the surface density of the young (left) and old (right) populations we use. In (a) we use a slice of $0.5<|z|<3\kpc$, and in (b) we use $|y|<5\kpc$. White dashed lines represent longitudes (a) and latitudes (b) from $-20\degrees$ to $20\degrees$ in steps of $5\degrees$ (straight lines) and distances from the Sun from $5.1$ to $11.1\kpc$ in steps of $1\kpc$ (curves). The $\Rgc=3.5\kpc$ limit is also shown, as a circle in panel (a).}
    \label{fig:density_youngold}
\end{figure*}

We define `young' and `old' stellar populations in the model as follows. The young population consists of all the stars in the age range ${4\mathrm{-}7\Gyr}$\footnote{In this way we differ from \citet{gough-kelly2022} who included all stars younger than $7\Gyr$ in their young population, and all stars older than 9 in their old population.}. This population hosts a strong bar and X-shaped bulge. We take stars in the age range ${9.5-10\Gyr}$, as the old population, which is spheroidal and boxy. The young and old populations constitute ${\sim}19.2\%$ ($1{,}395{,}563$) and ${\sim}23.6\%$ ($1{,}713{,}858$) of all bulge stars, respectively (computed over the full volume ${|l|<11\degrees}$, ${|b|<13\degrees}$ and ${\Rgc<3.5\kpc}$), as shown in Fig.~\ref{fig:cumu_age}. 

The model has a prominent nuclear stellar disc, where the youngest stars form \citep{Cole2014, Debattista2015}. This structure is confined to the galactic plane, with $|z|\lesssim 150\pc$ (after the coordinate re-scaling). In this study we are not interested in this structure. We avoid it by excluding stars younger than $4\Gyr$, and by only considering stars at $|z|>0.5\kpc$, which at the galactic centre corresponds to $|b|\gtrsim3.6 \degrees$. The resulting young and old populations are shown in Fig.~\ref{fig:density_youngold}. 


\section{Face-on kinematics}
\label{sec:faceon_maps}

In this section we explore the differences between the face-on kinematics of the strongly barred young population, and the weakly barred old population defined for the model in Section \ref{subsec:model-bar}. We bin the stars in $x$-$y$ space with $0.5<|z|/\kpc<3$.

In Section \ref{subsec:cylindrical_maps} we characterise the kinematics of the two populations using the intrinsic galactocentric cylindrical velocities $v_R$-$v_\phi$, while in Section \ref{subsec:galactic_xy_maps} we explore how these project into the heliocentric frame by using velocities in galactic coordinates, $v_r$-$v_l$. 
Mean velocity maps of similar populations, for the same model, have already been presented in \citet{gough-kelly2022}; here we also present mean velocity maps to aid with the interpretation of the different velocity ellipses.

\subsection{Galactocentric velocities}
\label{subsec:cylindrical_maps}

\begin{figure*}
    \centering
    \includegraphics[angle=0.,width=\hsize]{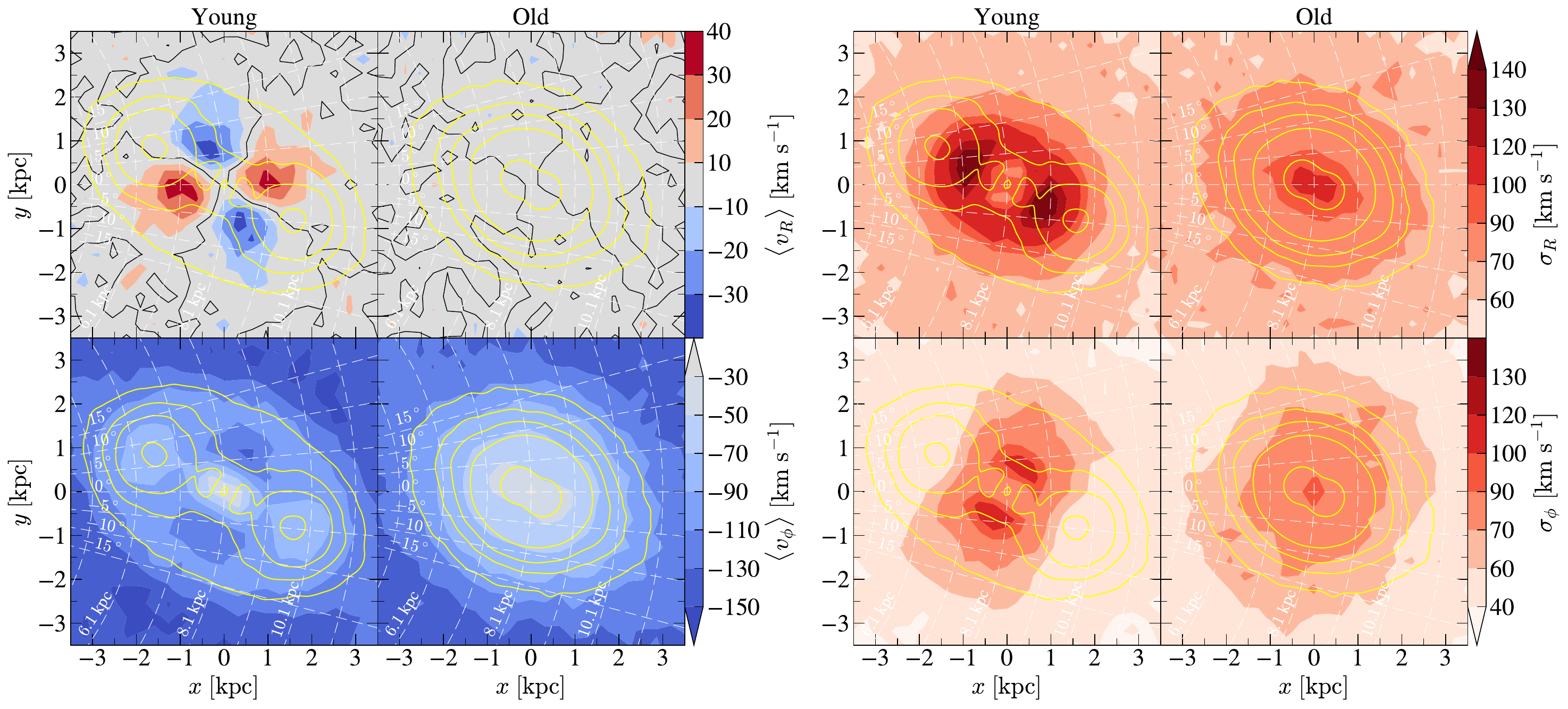}
    \caption{Face-on kinematic maps of the model using galactocentric cylindrical velocities $v_R$-$v_\phi$ (see Fig.~\ref{fig:xy_gal_vxvy} for $v_r$-$v_l$ instead). We show two blocks: mean velocities (left) and the corresponding dispersions (right). Each block contains two columns, corresponding to the young (left) and old (right) populations. Black solid contours follow values of $v_R = 0$. Yellow contours outline the density distribution (see Fig.~\ref{subfig:top_density_youngold}). The Sun is located at $(x,y) = (-8.1,0)\kpc$. White dashed lines represent longitudes from $-15\degrees$ to $15\degrees$ in steps of $5\degrees$ (straight lines) and distance from the Sun, $d$, from $6.1$ to $10.1\kpc$ in steps of $1\kpc$ (curves).}
    \label{fig:xy_cyl_vxvy}
\end{figure*}

Fig.~\ref{fig:xy_cyl_vxvy} shows maps of the average radial and tangential velocity and velocity dispersion of the galactocentric cylindrical components $v_R$ and $v_\phi$. The left block shows $\meanvR$ and $\meanvphi$, and the right shows their dispersions, $\sigma_R$ and $\sigma_\phi$. The left and right columns in each block correspond to the young and old populations respectively.

The mean velocities for the old population exhibit the expected radial and tangential velocity fields for a relatively slowly rotating, nearly axisymmetric component. The old population has $\meanvR \sim 0$ everywhere, and a relatively slow rotation, slightly elongated along the bar major axis, with $\meanvphi$ increasing slowly with $R$. On the other hand, the quadrupole pattern in $\meanvR$ for the young stars reveals the strongly barred nature of this population. We find a good match with the quadrupole found by \citet[][their figure 2]{bovy2019} in the MW. This pattern arises from the radial streaming along the bar.
The fast streaming motions of young stars along the bar produce $\meanvphi$ peaks along the bar's minor axis, ${\sim}1\kpc$ away from the centre. The mean velocity then drops in the regions of highest density (in the arms of the X-shape), indicative of stars reaching the apocentre of their orbits. These results are in agreement with those obtained by \citet[][their figure 5]{gough-kelly2022}.

The right block in Fig.~\ref{fig:xy_cyl_vxvy} shows the galactocentric velocity dispersions of the same populations. The peaks of high $\sigma_R$ for the young population are caused by the crossing of the high velocity streaming motions along the bar major axis. Weaker peaks are present in $\sigma_\phi$, on the bar's minor axis. The dispersions peak away from the centre of the galaxy.
The velocity dispersions of the old stars instead peak at the centre, and then decline slowly radially.

\begin{figure*}
    \centering
    \includegraphics[width=\textwidth]{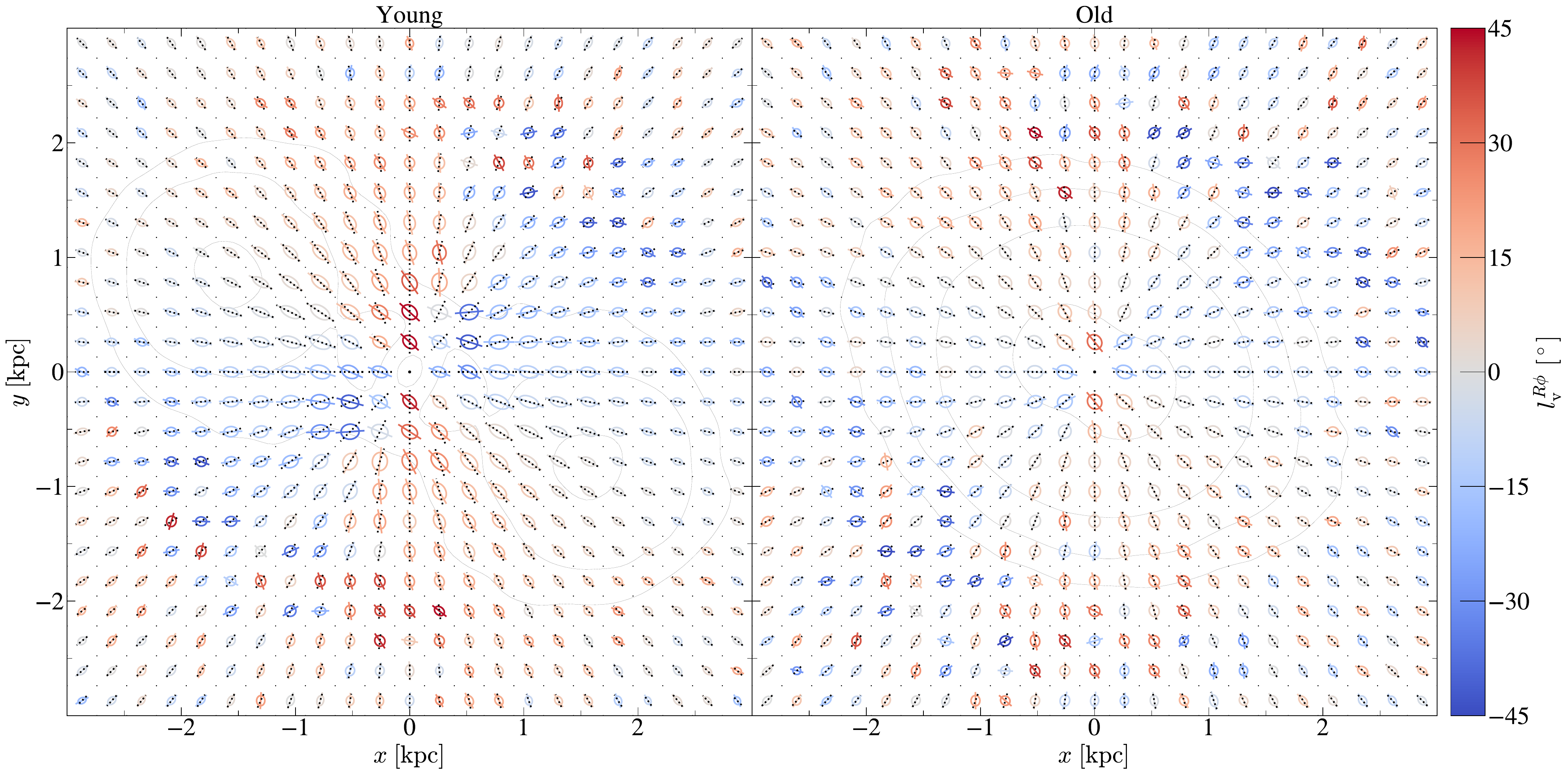}
    \caption{Face-on view of the velocity ellipses of the young (left) and old (right) stars in the model, coloured by the vertex deviation $\vertexabsRphi$. We have overlaid on each ellipse both its semi-major axis (solid coloured line) and a dotted black line pointing in the radial direction. The black contours outline the density distribution. The grid of black points delimits the different $x$-$y$ bins. All bins contain at least 50 stars. We omitted the central ellipse due to the instability of the galactocentric unit vectors there.}
    \label{fig:xy_ellipses}
\end{figure*}

Fig.~\ref{fig:xy_ellipses} presents the velocity ellipses for the young (left) and old (right) populations. Even though the ellipses are defined in velocity space, it is useful to overlay them in position space because the orientation of their semi-major axis points in the (local) direction of highest dispersion in the $x$-$y$ plane. We have centered the ellipses on the centre of each bin (in velocity space they would be offset from the origin by the mean velocity values). 
The ellipses are colour-coded by the vertex deviation, $\vertexabsRphi$. Positive or negative values indicate the semi-major axes are tilted anti-clockwise or clockwise, respectively, with respect to the local galactocentric radial vector, $\Rhat$. An overall radial alignment of the old population is evident, which results in a low vertex deviation everywhere within $R<1.5\kpc$ (ignoring the outer noisy regions with fewer stars), as expected for a nearly axisymmetric system. For the young stars, the ellipses within the bar are aligned along its semi-major axis, which causes radial alignment at the X-shape overdensities but azimuthal alignment along the bar's semi-minor axis within $R\lesssim1\kpc$. In the transition from radial to azimuthal alignment we find velocity ellipses with very strong tilts, reaching ${\vertexabsRphi \sim \pm45\degrees}$, a clear indication of the strong bar in this population.
\begin{figure*}
    \centering
    \includegraphics[width=\hsize]{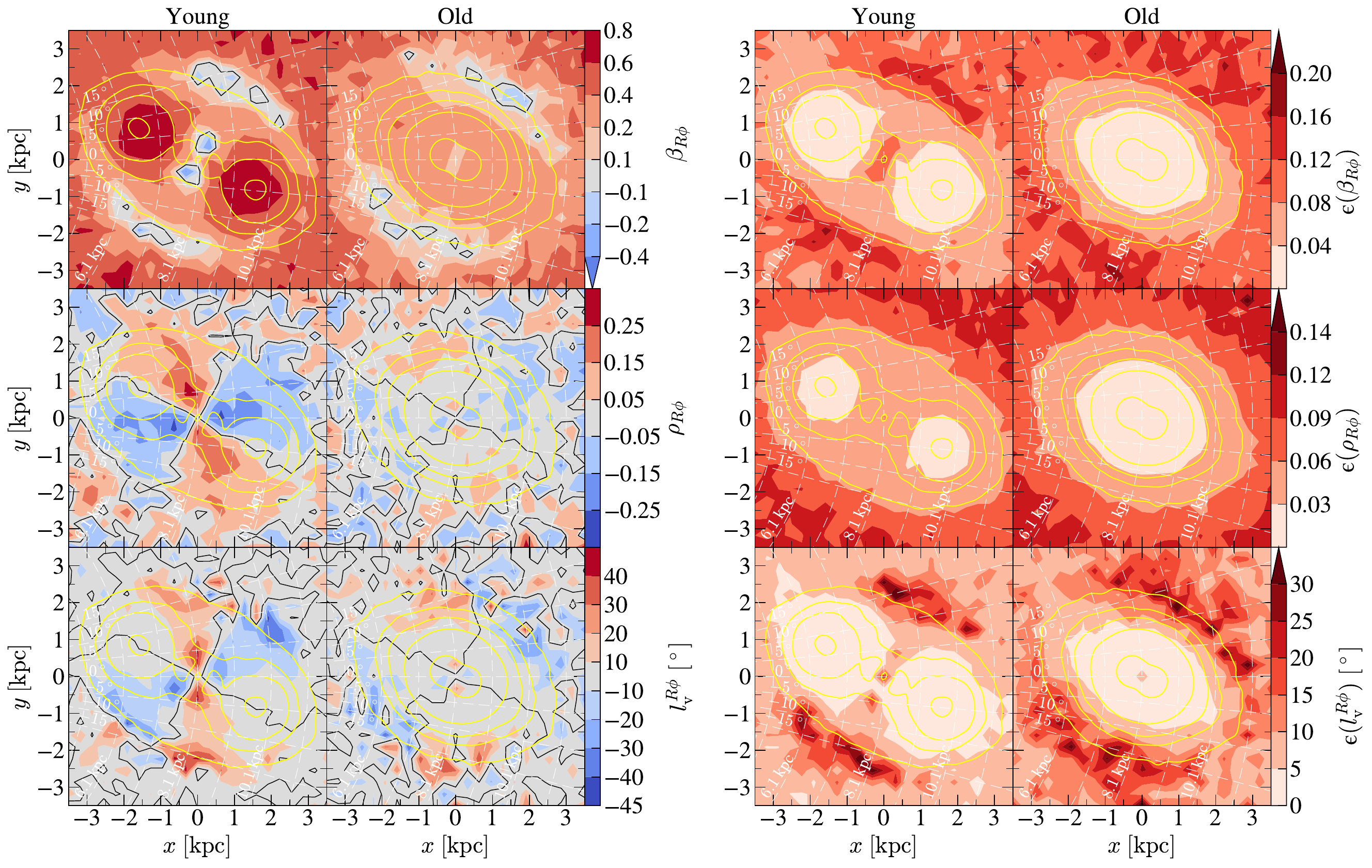}
    \caption{Left block: anisotropy, correlation and vertex deviation (top to bottom). Right block: the associated errors, computed using bootstrapping with 500 repetitions. Each block contains two columns, corresponding to the young (left) and old (right) populations. Black solid contours follow values of zero. Yellow contours outline the density distribution. The Sun is located at $(x,y) = (-8.1,0)\kpc$. White dashed lines represent longitudes from $-15\degrees$ to $15\degrees$ in steps of $5\degrees$ (straight lines) and distance from the Sun, $d$, from $6.1$ to $10.1\kpc$ in steps of $1\kpc$ (curves).}
    \label{fig:xy_cyl_anicorr}
\end{figure*}

Fig.~\ref{fig:xy_cyl_anicorr} quantifies the velocity ellipses across the inner galaxy for the two populations. The first row shows the anisotropy. The dispersion quadrupoles in Fig.~\ref{fig:xy_cyl_vxvy} imprint quadrupoles in the anisotropy of the young stars. The correlation of the young stars, seen in the middle row, shows a similar quadrupole to that in $\meanvR$, although with opposite signs. A quadrupole in $\vertexabsRphi$ results, which is particularly strong ($\pm45\degrees$) just outside the centre. The old population instead has mostly positive anisotropy everywhere, as expected for a nearly axisymmetric system. Its weak bar imprints weaker quadrupoles in the anisotropy, correlation and the vertex deviation compared with the young populations.

We estimate uncertainties using the bootstrap method. Given a sample of $N$ stars, after calculating the value of a quantity of interest, $x$, we obtain $B$ random samples with replacement, of the same size $N$, and compute the value of the same quantity, $x^*_i$, for each bootstrap sample $i$. We then estimate the uncertainty in $x$ as
\begin{equation}
    \upepsilon(x) = \sqrt{\frac{1}{B}\sum_i^k\left(x^*_i-x \right)^2}.
    \label{eq:bootstrap}
\end{equation}
We use $B=500$ bootstrap iterations.

The uncertainties in Fig.~\ref{fig:xy_cyl_anicorr} are high where the number of stars is small. The velocity ellipses of old stars have small uncertainties throughout the bulge region because this region is well populated. We also find small errors at the X-shape overdensities of the young stars. The anisotropy of the young stars has a relatively large error (${\sim}0.1$-$0.2$) along the bar's outer minor axis, approximately at the same location as the negative $\anicyl$ lobes. The correlation quadrupole has $|\corrcyl|>0.15$ while the errors are $\upepsilon(\corrcyl)<0.06$. The correlation quadrupole is therefore a robust feature of non-axisymmetry. The regions of $\upepsilon(\vertexabsRphi)\gtrsim10\degrees$ along the bar's minor axis correspond to ellipses with very small $\anicyl$ and $\corrcyl$. Nearly isotropic, weakly correlated populations have velocity ellipses with ill-defined major axis directions so their vertex deviations are unreliable. The regions of the vertex deviation quadrupole in Fig.~\ref{fig:xy_cyl_anicorr} have small associated errors, so this is also a robust feature of non-axisymmetry.

\subsection{Velocities in the heliocentric reference frame}
\label{subsec:galactic_xy_maps}

\begin{figure*}
    \centering
    \includegraphics[width=\hsize]{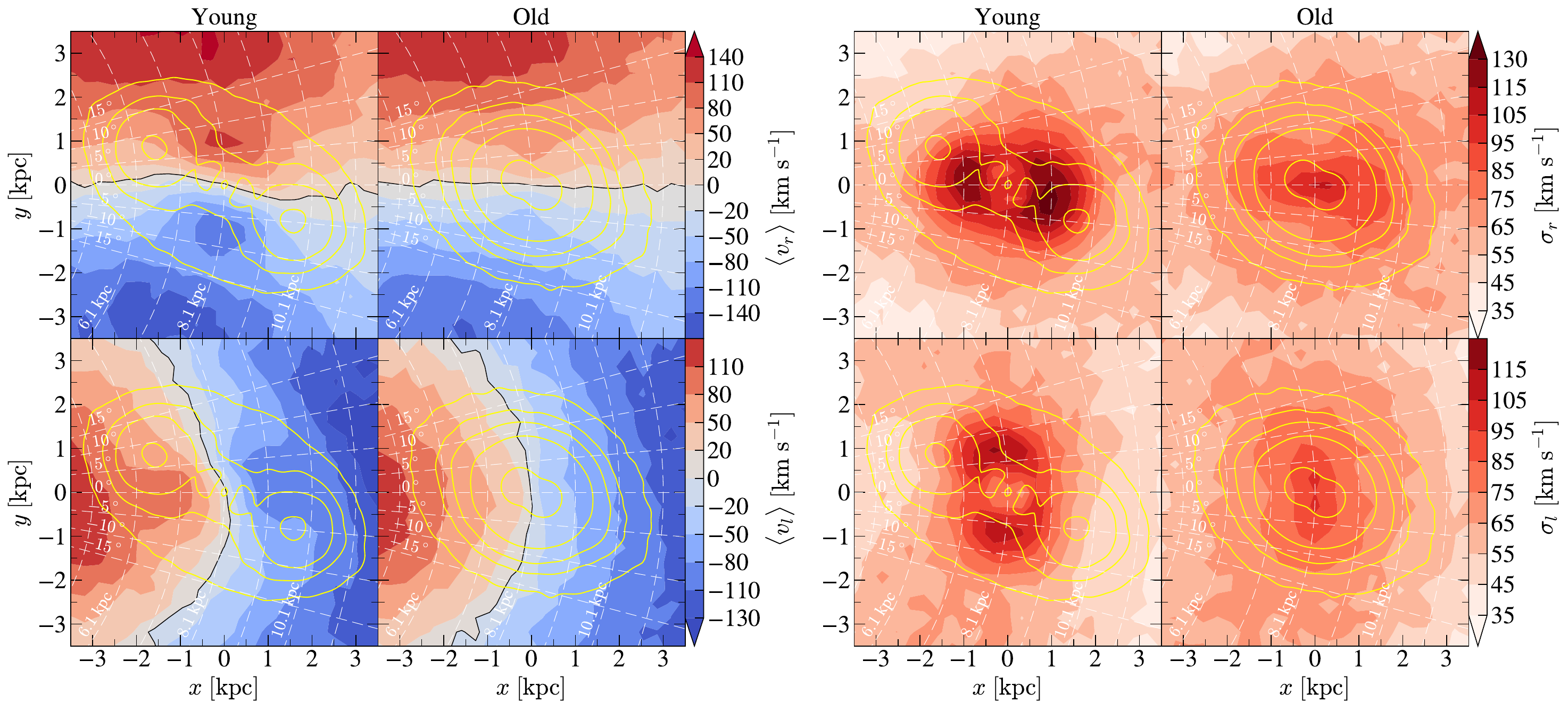}
    \caption{Face-on kinematic maps of the heliocentric galactic velocities, $v_r$ and $v_l$, for the model (see Fig.~\ref{fig:xy_cyl_vxvy} for $v_R$-$v_\phi$ instead). We show two blocks: mean velocities (left) and the corresponding dispersions (right). Each block contains two columns, corresponding to the young (left) and old (right) populations. Black solid contours follow values of zero. Yellow contours outline the density distribution (see Fig.~\ref{subfig:top_density_youngold}). The Sun is located at $(x,y) = (-8.1,0)\kpc$. White dashed lines represent longitudes from $-15\degrees$ to $15\degrees$ in steps of $5\degrees$ (straight lines) and distances away from the Sun from $6.1$ to $10.1\kpc$ in steps of $1\kpc$ (curves).}
    \label{fig:xy_gal_vxvy}
\end{figure*}

We now turn to the bulge as seen from the Sun by considering kinematics in Galactic coordinates. Fig.~\ref{fig:xy_gal_vxvy} is the equivalent of Fig.~\ref{fig:xy_cyl_vxvy} in the heliocentric Galactic coordinates; it shows $\meanvr$, $\meanvl$, $\sigma_r$ and $\sigma_l$. 
The velocities of old stars are nearly symmetric between positive and negative longitudes, which are a result of the near axisymmetry of this population. In contrast, the streaming motions along the bar of the young stars cause the near and far regions at small longitudes to have $\meanvr$ of the opposite sign to $l$. This twist in the velocities with respect to the lines of $l$ gives rise to \textit{forbidden velocities} \citep{gough-kelly2022} and constitutes part of the observable kinematic imprint of the bar. The strong bar in the young population also causes a slight misalignment between the $\meanvl=0$ line and the line of constant $d=8.1\kpc$.

The velocity dispersion quadrupoles are no longer aligned along the bar's principal axes when heliocentric velocities are used. Projection results in a weaker quadrupole for young stars in $\sigma_r$ than in $\sigma_R$, while the quadrupole in $\sigma_l$ is significantly stronger than in $\sigma_\phi$; both quadrupoles are clearly distorted by the effect of perspective.

\begin{figure*}
    \centering
    \includegraphics[width=\hsize]{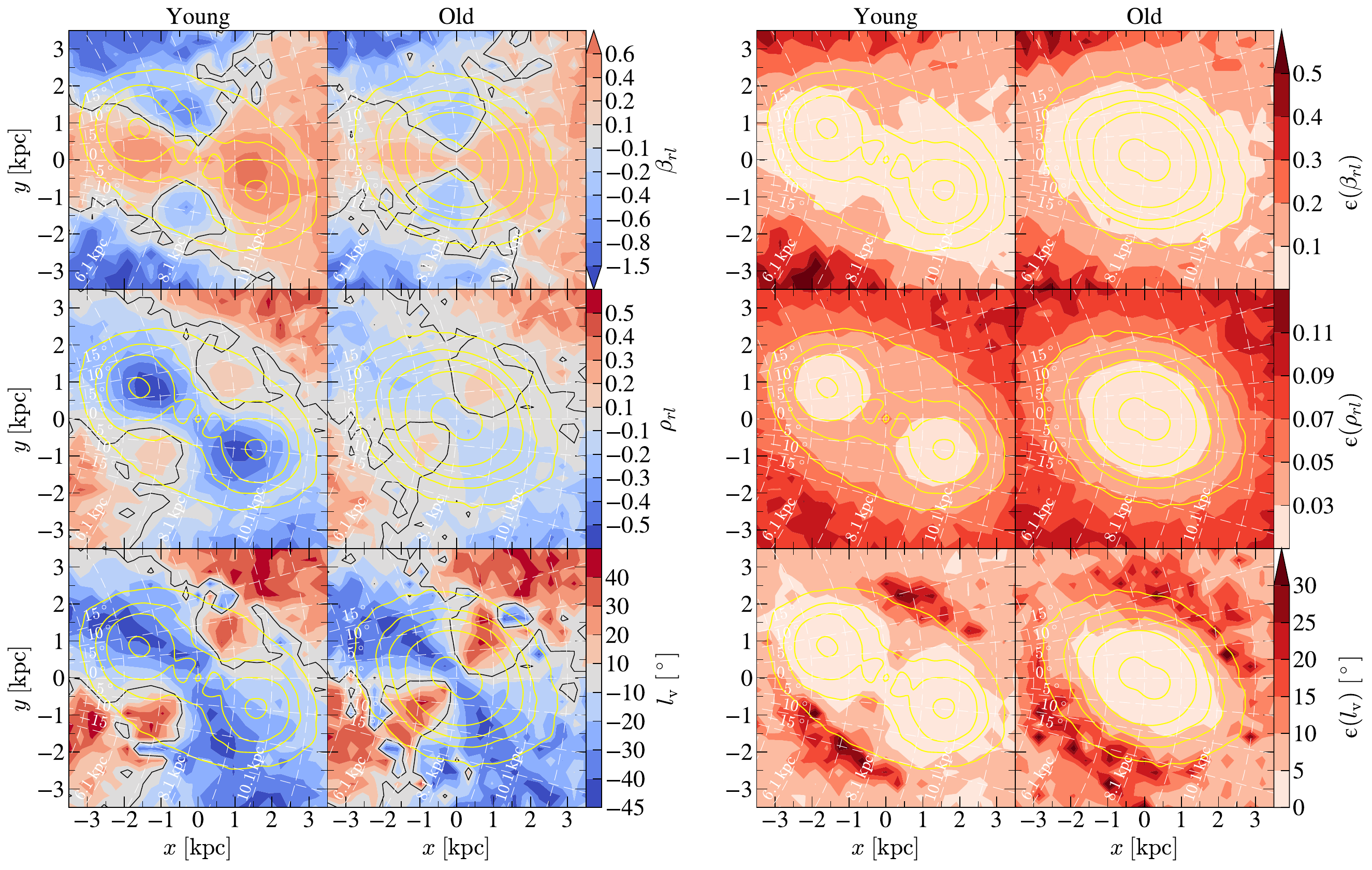}
    \caption{Same as Fig.~\ref{fig:xy_cyl_anicorr}
    but now showing, from top to bottom, anisotropy ($\ani$), correlation ($\corr$), and the vertex deviation, $\vertexabs$. }
    \label{fig:xy_gal_anicorr}
\end{figure*}

Fig.~\ref{fig:xy_gal_anicorr} shows a strong quadrupole in $\ani$ for the young population. The old stars present a weaker quadrupole which arises largely from projecting their direction of highest dispersion, $\Rhat$, onto the Galactic coordinates. The same geometric projection gives rise to a quadrupole in the correlation and the vertex deviation of old stars, rotated by $45\degrees$ from that of the anisotropy. In a perfectly axisymmetric population, the black solid contours of $\ani=0$, $\corr=0$ and $\vertexabs=0\degrees$ would join at the origin. This does not occur due to the weak bar of the old population. The effect is more extreme for the young stars, with the positive poles of their $\corr$ and $\vertexabs$ quadrupoles showing a wider separation. Moreover, the correlation of the young population is significantly stronger.

In the central kpc along $l=0\degrees$, the vertex deviation is somewhere between $-20\degrees$ and $-30\degrees$. This is not surprising as the bar angle is $-27\degrees$, and the direction of highest dispersion is closely related to the bar direction at $l=0\degrees$ due to the streaming motions along the bar. This is what enabled \citet{simion2021} to estimate the bar's angle.

The right block of Fig.~\ref{fig:xy_gal_anicorr} shows the bootstrap errors on the heliocentric anisotropy, correlation, and vertex deviation. The old stars have small errors everywhere in the bulge region, just as in Fig.~\ref{fig:xy_cyl_anicorr}. For the young stars, the anisotropy error is smaller ($<0.1$) along the bar's minor axis than for the intrinsic velocities in Fig.~\ref{fig:xy_cyl_anicorr}. The errors on \corr\ and \vertexabs\ look similar to those in Fig.~\ref{fig:xy_cyl_anicorr}: ${\upepsilon(\corr)<0.07}$ and ${\upepsilon(\vertexabs)\lesssim10\degrees}$ across most of the bulge region, and larger ${\upepsilon(\vertexabs)}$ peaks located at points of small $\ani$ and $\corr$, along the bar's minor axis. Therefore, the negative correlation and vertex deviation bands along $l=0\degrees$ in Fig.~\ref{fig:xy_gal_anicorr} are stable features of non-axisymmetry.

Thus the vertex deviation of stars along the $l=0\degrees$ line is a useful probe of the bar's strength even when the line-of-sight distance of tracer populations is uncertain. The vertex deviation along this direction is robust because $\vrhat$ and $\vlhat$ (and consequently $\vertexabs$) are well-defined and stable for the whole extent of the bulge.


\section{Observational data}
\label{sec:data}

In order to test the model's predictions, we define an observational sample of stars in the MW's bulge. In the observations we use metallicity as a proxy for age, with the expectation that older stars are more metal-poor.

\subsection{APOGEE and \textit{Gaia} DR3}
\label{subsec:apogee-gaia}

We use data from the Apache Point Observatory Galactic Evolution Experiment (APOGEE) survey \citep{Majewski_2017}, part of the Sloan Digital Sky Survey (SDSS). We use APOGEE-1 and APOGEE-2 data, from SDSS-III and -IV respectively \citep{eisenstein2011, blanton2017}. APOGEE is a high-resolution spectroscopic survey that observes in the near-infrared \textit{H} band ($1.51$-$\SI{1.70}{\micro \meter}$). These wavelengths allow it to collect detailed chemical and kinematic information from the inner Galaxy, piercing through the dust present in the disc. Its main targets are red giant stars from the red-giant branch (RGB), the asymptotic giant branch (AGB) and the red clump (RC), which are luminous tracers present in most stellar populations \citep{Majewski_2013,Zasowski2017}. 

We use the cleaned bulge dataset from \citet{rojas_arriagada_2020}, comprised of $13{,}031$ APOGEE DR16 stars within ${|l|<11\degrees}$, ${|b|<13\degrees}$ and ${R_\mathrm{GC}<3.5\kpc}$. We use the radial velocity of the stars computed from the spectroscopic data using the pipeline of \citet{Nidever2015}, and the metallicities, \feh, computed using the APOGEE Stellar Parameters and Chemical Abundances Pipeline (ASPCAP) in \citet{GarciaPerez2016}. We also use the stellar distances computed using the spectro-photometric method in \citeauthor{rojas-arriagada2017} (\citeyear{rojas-arriagada2017,rojas-arriagada2019}) with $JHK_\mathrm{s}$ photometric values from 2MASS \citep{Skrutskie_2006}. Their method infers distances and line-of-sight reddening simultaneously through isochrone fitting, using PARSEC isochrones\footnote{Available at \url{http://stev.oapd.inaf.it/cgi-bin/cmd}} covering ages from 1 to 13~Gyr and metallicities from $-2.2$ to $+0.5$~dex. Each star is placed in $T_{\mathrm{eff}}$-$\log g$-metallicity space, and the normalized distance $d_{\mathrm{iso}}$ to each model star is computed, accounting for observational errors. Two physical weights are applied: $W_{\mathrm{es}}$, proportional to the mass difference between consecutive model stars along the isochrone, which corrects for oversampling in short-lived phases and assigns greater weight to longer evolutionary stages, and $W_{\mathrm{IMF}}$, proportional to the relative number of stars in each mass interval according to the initial mass function. The total weight assigned to each model star $j$ is then given by
$$
W_j = W_{\mathrm{es}} W_{\mathrm{IMF}} \exp(-d_{\mathrm{iso}}).
$$
Weighted estimates of the theoretical absolute magnitudes ($M_J$, $M_H$, $M_{K_s}$) are derived, and by comparing them with the observed 2MASS magnitudes, the distance and reddening for each star are determined.

We tested our distances against those from the \citet[][2010 edition]{harris1996} catalogue\footnote{Available at \url{https://physics.mcmaster.ca/~harris/mwgc.dat}} for a sample of globular cluster stars observed by APOGEE, and against \textit{Gaia} DR3 Bayesian distances \citep{bailerjones2021} for a sample of nearby ($d_\mathrm{CBJ}<3\kpc$) APOGEE RGB/AGB stars, selected to be representative of the sample analysed in this work, and whose \textit{Gaia} astrometry have relative parallax errors smaller than $5\%$. Taking as distance errors the absolute differences between the spectro-photometric and reference distances, we find a median fractional error of $8.3\%$, with the $25$th and $75$th percentiles at $3.6\%$ and $16.7\%$ respectively. Comparing these external errors with the internal errors derived from the isochrone fitting, which indicate the quality of the fit, we find a moderate correlation (Pearson $\rho=0.52$) and that the internal error underestimates the external error, which indicates that the internal error does not encapsulate all sources of error. This is not an issue because the bootstrap error is the dominant source of error in our results even if we assume a constant fractional distance error of $20\%$. We analyse the effect of even larger distance uncertainties in Section \ref{sec:distance_error}.

\citet{rojas_arriagada_2020} adopted proper motions for the $13{,}031$ star sample from \textit{Gaia} DR2 \citep{gaiaDR2}. In this study we update the proper motions with \textit{Gaia} DR3 \citep{gaiaDR3}. Of the $13{,}031$ stars, $341$ do not have a match in \textit{Gaia} DR3. Conversely $4{,}395$ DR2 stars have two or more matches in DR3, and we select as best match the source with the smallest absolute difference in G magnitude. Leaving out 17 problematic sources where the magnitude difference is larger than 1 mag, we are left with $12{,}673$ valid matches. From these, 716 have renormalised unit weight error (RUWE) larger than 1.4, and we removed them. Of the remaining stars, $1{,}422$ and $282$ are missing $\feh$ and proper motions, respectively, and we remove them too, leaving us with $10{,}486$ stars.

\subsection{The bulge sample}

We stick to the ${|l|<11\degrees}$, ${|b|<13\degrees}$ and ${\Rgc<3.5\kpc}$ selections applied by \citet{rojas_arriagada_2020}. We assume a Sun - Galactic Centre distance of $R_0=8.1\kpc$ \citep{gravity2018,Bobylev2021}, and adopt a Galactocentric Solar velocity ${(U,V,W)_\odot=(12.9,245.6,7.78)\kms}$ \citep{Reid_2004,Drimmel_2018,gravity2018}. We convert all our velocities to the Galactic Standard of Rest (GSR) using the \texttt{PYTHON} package \texttt{galpy} \citep{galpy}.

The metallicity values range from $-2$ to $0.6$~dex. We exclude stars with $\feh<-1 \dex$, which amount to $548$ stars (${\sim}5.2\%$), since very metal-poor stars are dominated by the halo (see \citealt{madeline2021}), which we are not interested in for this study. 
After these selections, our final bulge sample contains $9{,}884$ stars.

\begin{figure*}
    \centering
    \includegraphics[width=0.75\hsize]{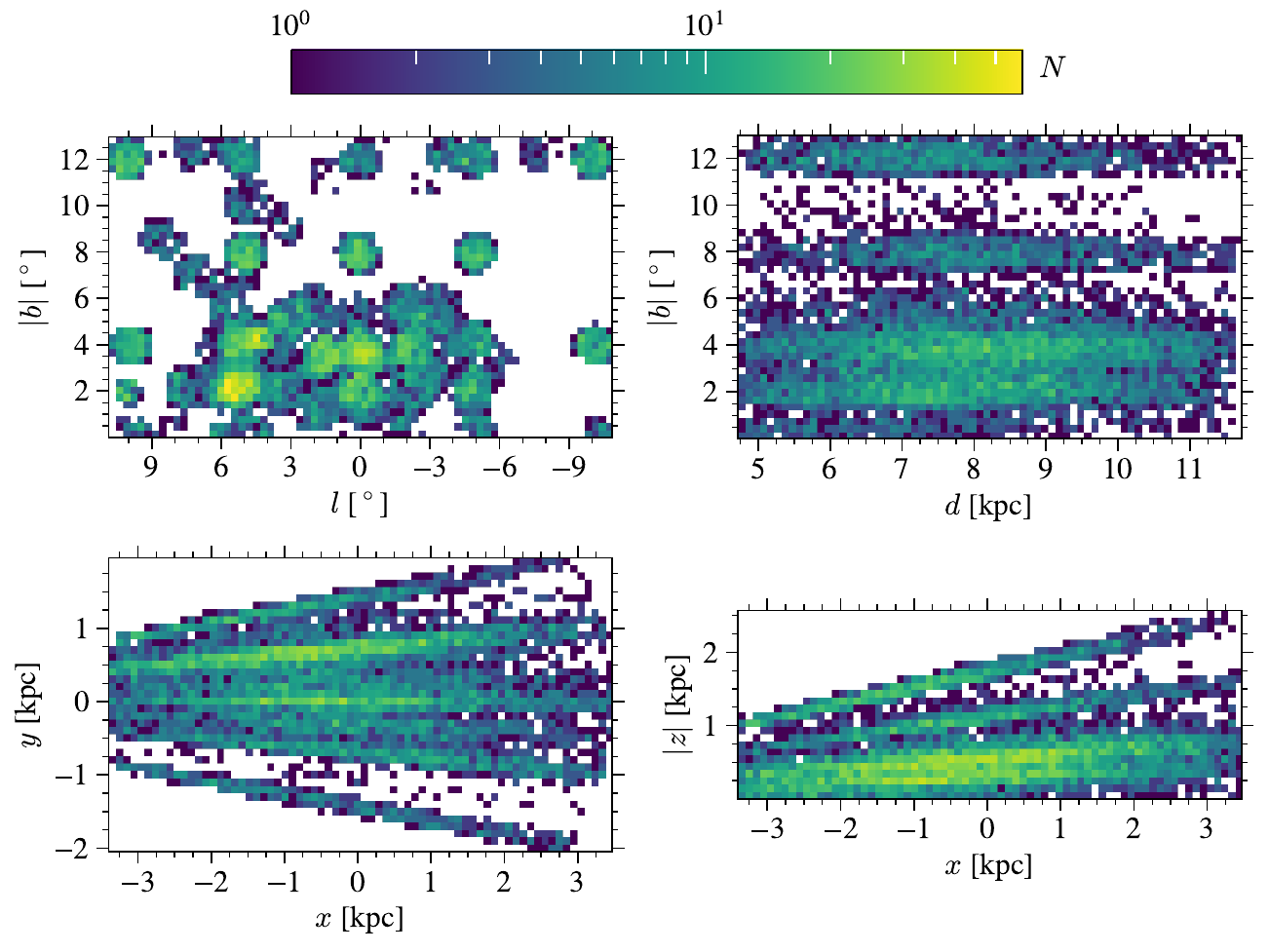}
    \caption{Distribution of the APOGEE stars in our bulge sample in different spatial representations. The top row shows the distribution in Galactic coordinates, while the bottom row shows Cartesian Galactocentric coordinates. The Sun is located at $(x,y)_\odot=(-8.1,0)\kpc$. All the heatmaps are shown on a log scale, with white indicating bins with no stars.}
    \label{fig:apogee_windows}
\end{figure*}

We assume that the bulge of the MW, and that of the model, are symmetric with respect to the mid-plane \citep{Wegg2015}. Thus we follow \citet{gough-kelly2022} and \citet{rojas_arriagada_2020} in folding the stars across the mid-plane, by projecting ${z'=-z}$ and ${v_z'=-v_z}$ for ${z<0}$. This allows us to increase the number of stars in each spatial bin, which is particularly useful given the relatively small number of stars in the observational data. Hereafter we work with $|b|$ and $|z|$. Fig.~\ref{fig:apogee_windows} shows the spatial distribution of the $9{,}884$ stars in our bulge sample. The top row shows the view in Galactic coordinates, and the bottom row in Galactocentric Cartesian coordinates. In the top row we can see that the majority of stars are located at relatively low latitudes, $|b|<5\degrees$.
Most stars were observed by APOGEE in fields of view of ${\sim}2\degrees$ in diameter, which reach deep into the bulge. We note our APOGEE sample may be biased in distance for different metallicities, as previous studies have suggested that APOGEE star selection is skewed towards metal-poor stars at greater distances, particularly behind the Galactic Center (\citealt{queiroz2021}).

\begin{figure}
    \centering
    \includegraphics[width=\columnwidth]{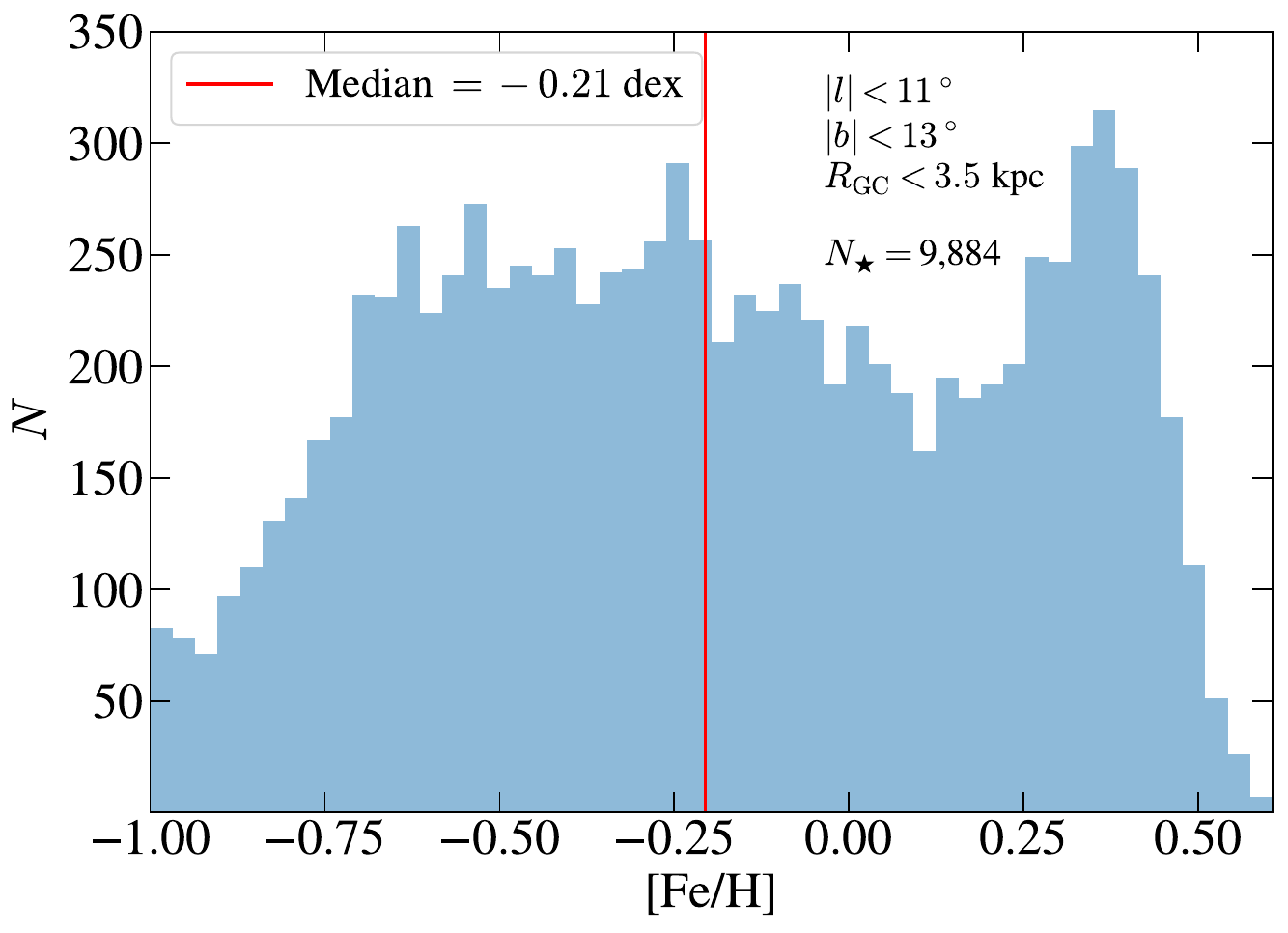}
    \caption{Metallicity distribution function of the APOGEE stars in the bulge region. The vertical line at $-0.21$ indicates the median metallicity.}
    \label{fig:metallicity}
\end{figure}

The metallicity distribution of the bulge stars is shown in Fig.~\ref{fig:metallicity}. There is a clear peak at ${\sim}0.4\dex$, consistent with component A identified by \citet[][their figure 12]{ness2013a}, while the rest of the distribution forms a broader cluster around ${\sim}-0.35\dex$, consistent with an aggregate of components B and C in \citet{ness2013a}. The median metallicity is $-0.21\dex$. In the kinematic analyses that follow, we define a metal-poor and metal-rich population by splitting the distribution around the median metallicity. This leaves enough stars to perform a broad qualitative comparison with the results from the old and young populations in the model, respectively.

We use a Monte Carlo propagation to estimate the effect of individual measurement (distance\footnote{We propagated the internal distance error which, as discussed in Section~\ref{subsec:apogee-gaia}, is only one source of error (isochrone fitting quality). This is the best we can do as it is the only source of errors which we can quantify on a star-by-star basis.} and proper motions) uncertainties on the velocity ellipse properties for the observational data. We find that the resulting error is negligible relative to the bootstrap error. Therefore, we ignore the measurement uncertainties and, as for the model, we use $B=500$ bootstrap iterations following Eqn.~\ref{eq:bootstrap} to estimate the errors on the observational data kinematics.


\section{Velocity ellipses along the minor axis}
\label{sec:minor_axis_comparison}

We now compare the velocity ellipses of the model, and of the APOGEE data, as seen from the Sun (\ie\ using heliocentric $v_r$-$v_l$ velocities) along the bulge minor axis, $|l|<2\degrees$.

\subsection{Variation with latitude}
\label{subsec:latitude}

A significant fraction of APOGEE stars is close to $l=0\degrees$, as can be seen in the top left panel of Fig.~\ref{fig:apogee_windows}. This allows us to explore the variation of the vertex deviation with latitude.  
We select stars at ${|l|<2\degrees}$, ${1.5\degrees<|b|<13\degrees}$ and ${\Rgc<3.5\kpc}$. We exclude the ${|b|<1.5\degrees}$ region to avoid the potential influence of the nuclear stellar disc. This is a generous cut for the MW (see \citealt{sormani2022}), but required by the larger nuclear disc in the model (see Section \ref{subsec:model-bar}). In every panel we show in a lighter colour the results using a reduced radial cut of ${\Rgc<2\kpc}$.

Given the latitude distribution of the APOGEE stars (see Fig.~\ref{fig:apogee_windows}), we divide the ${1.5\degrees<|b|<7\degrees}$ region into 2 bins, each containing an equal number of stars. Additionally, we treat the pencil-beam fields of view centered at ${|b|\sim8\degrees}$ and ${12\degrees}$ as separate bins, resulting in a total of 4 bins. For the model, instead, we divide the latitude range into 8 equally-spaced bins.
Since the simulation is not a detailed model of the MW, we are only interested in comparing the two datasets qualitatively, so this binning suffices.
\begin{figure}
    \centering
    \includegraphics[width=0.9\columnwidth]{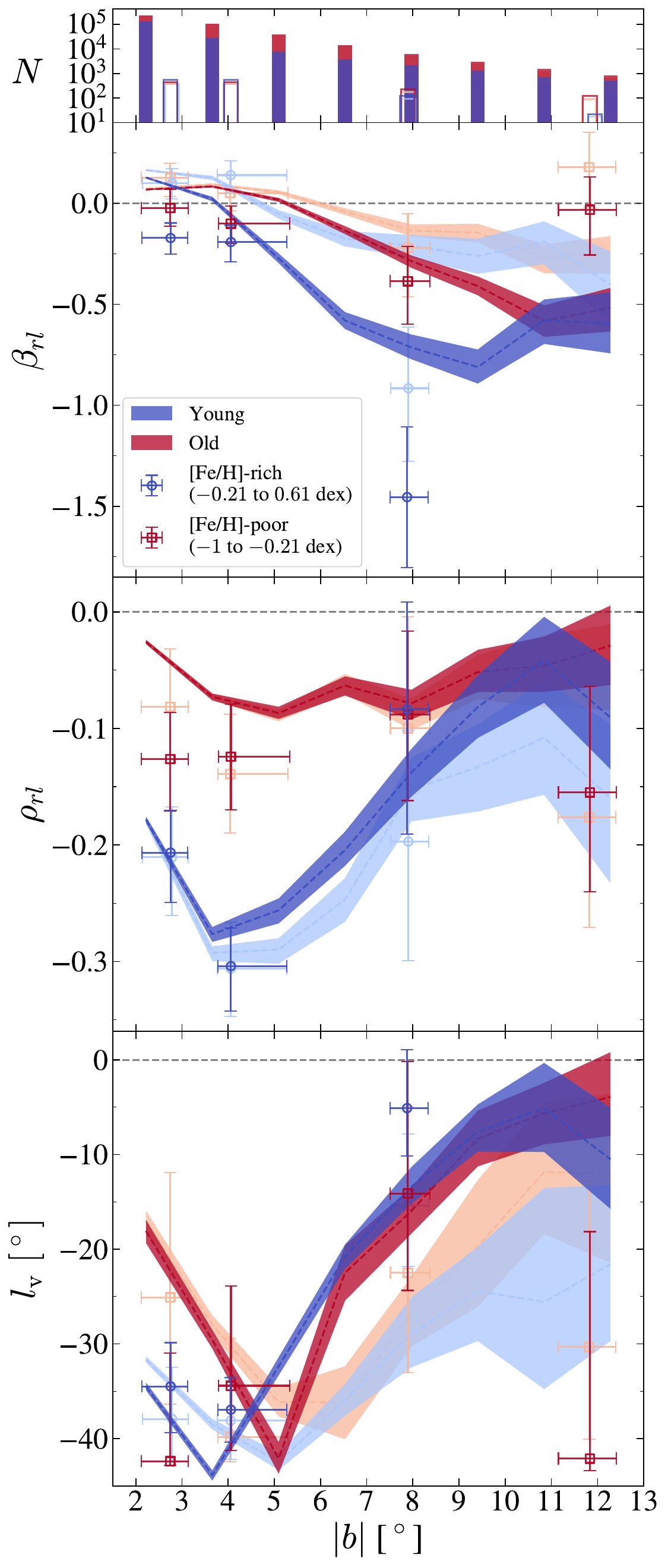}
    \caption{Vertical profiles of anisotropy (top), correlation (middle) and vertex deviation (bottom) along the bulge minor axis, $|l|<2\degrees$, within ${\Rgc<3.5\kpc}$ (darker) and ${\Rgc<2\kpc}$ (lighter). The shaded areas show the model while the data points show the APOGEE data. The number of stars in each bin is shown in the bar plot at the top, with the filled histograms showing the model and the open ones showing the APOGEE data. Only data points with at least 50 stars are shown; there are less than this metal-rich stars in the highest $|b|$ bin, so this bin is excluded.}
    \label{fig:kinematics_lat}
\end{figure}
Fig.~\ref{fig:kinematics_lat} shows the results for the anisotropy (top), correlation (middle), and vertex deviation (bottom) as a function of $|b|$. We represent the model by shaded areas, with the width at each point measuring the $68\%$ bootstrap confidence interval. The observational data are shown as points, placed at the median latitude of the stars in each bin, with kinematic error bars measuring the $68\%$ bootstrap confidence interval, and position error bars measuring half the distance between the point and the bin's edges.

Given the relatively strong positive anisotropy, $\ani$, of the model in the $x$-$y$ plane along $l\sim0\degrees$ across the full depth of the bulge, particularly for the young stars, as seen in the top row of Fig.~\ref{fig:xy_gal_anicorr}, we might expect to find strong $\ani>0$ in the top panel of Fig.~\ref{fig:kinematics_lat}. However, the opposite signs of $v_l$ on the near and far sides, seen in Fig.~\ref{fig:xy_gal_vxvy}, lead to a $\sigma_l$ almost as large as $\sigma_r$ at low latitudes, and even larger at higher latitudes. As a result, Fig.~\ref{fig:kinematics_lat} shows that the anisotropy of both populations in the model starts small and positive at low latitudes, then crosses the isotropy line (at $|b|\sim3.7\degrees$ and ${\sim}5.3\degrees$ for the young and old populations, respectively), grows stronger and negative beyond that, with a steeper gradient for the young stars, and eventually plateaus at the highest latitudes. 
The anisotropy of the APOGEE data follows a similar trend, with isotropic values near the plane, growing negative away from it. An exception is the metal-poor data point at the highest latitude, which returns to isotropy. This behaviour might be influenced by contamination from the thick disc (\citealt{rojas_arriagada_2020} found a metallicity distribution consistent with the thick disc at $|b|>10\degrees$). More data at higher latitudes would help to ascertain the behaviour there.
In the model, the effect of changing the radial range to $\Rgc < 2\kpc$, shown in lighter colour, is a shift of the curves towards $\ani = 0$, which is more prominent for the young stars away from the plane. The resulting effect is a flattening of the curves. The anisotropy responds to the change in radius in part because of the broadening caused by the opposite signs of the near and far sides of the galactic rotation, which itself varies radially. For the young stars the anisotropy changes also due to \textit{forbidden velocities} (see Section \ref{subsec:galactic_xy_maps}) becoming more important (as they exist within the inner kpc); the cancellation of the positive and negative sides then leads to a larger $\sigma_r$. For the old stars it is due to $\sigma_l$ decreasing, given $v_l$ increases radially, so the rotational broadening is reduced. The effect is not as strong close to the plane for both populations because their density is dominated by the hot dense region within $\Rgc\lesssim1\kpc$, as shown in Fig.~\ref{subfig:side_density_youngold}. The dominance of this inner region reaches further away from the plane for the old stars. 
The effect of reducing the radial range on the APOGEE data is also an upwards shift of the curve.

The correlation, shown in the middle panel of Fig.~\ref{fig:kinematics_lat}, is negative for all populations, in both the model and the APOGEE data. The correlation of the young stars in the model and metal-rich ones in the observations exhibit a U-shaped profile at $|b|<9\degrees$, with the largest (most negative) values in the interval $3\degrees<|b|<6\degrees$, dropping in magnitude above this. The correlation for the old and metal-poor populations is also negative but smaller in magnitude and flatter, showing less variation with latitude. The effect of the radial change is small to negligible in the correlation, for both the model and the APOGEE stars (except for the metal-rich bin at $|b|\sim8\degrees$, which, however, has very large error bars). The correlation of the young stars in the model, unlike the old ones, does experience a small ($\lesssim0.05$) systematic shift towards stronger values except at the lowest latitude.

The bottom panel of Fig.~\ref{fig:kinematics_lat} shows the vertical profile of the vertex deviation, $\vertexabs$. Both the young and old populations in the model exhibit a V-shape below $|b|<7\degrees$, with magnitude peaks of equal strength (reaching nearly $-45\degrees$). These peaks are located at the points nearest to the $\ani=0$ line (recall that a perfectly isotropic system with negative correlation would have $\vertexabs=-45\degrees$ exactly). Therefore, despite the weaker bar of the old stars, the vertex deviation is robustly capturing its non-axisymmetry, but not the fact that it is weaker than the bar in the young stars. We find similar trends for the metal-rich and metal-poor populations in APOGEE at ${|b|<7\degrees}$. The strong correlation and vertex deviation of the metal-rich stars are a clear signature of its non-axisymmetry. As for the model's old population, the near-isotropy of the metal-poor APOGEE stars leads to their vertex deviation revealing their non-axisymmetry, despite their relatively weak correlation. At the latitude of Baade's Window, $|b| \simeq 4\degrees$, both metal-rich and metal-poor populations have large (negative) \vertexabs\ values, as is seen in the MW for populations with $\feh > -0.7 \dex$ \citep{Soto_2007,Babusiaux2010}. 
Beyond ${|b|>7\degrees}$ the vertex deviation of all populations diminishes in magnitude, in part because of projection effects. The vertex deviation of the metal-poor stars at the highest latitude is strong but its error is very large. Again a larger sample at higher latitudes would help to elucidate the behaviour there.
Using the reduced radial range, $\Rgc < 2\kpc$, shifts the location of the peak $\vertexabs$ for the young stars, and reduces its amplitude below $|b| = 6\degrees$ in the old stars. In both populations, the amplitude of $\vertexabs$ increases substantially for $|b| > 6\degrees$.
There are hints of similar behaviour in the APOGEE data (\eg\ $|\vertexabs|$ increases for both metal-rich and metal-poor populations at $|b|=8\degrees$) but more data are required for the results to be significant.

Overall, we find that, at latitudes $|b|<7\degrees$, while vertex deviation serves as a blunt tool to detect non-axisymmetry, correlation works best at separating the young and old populations due to their differing bar strengths. This holds for the observations remarkably well too despite their reduced number of stars relative to the model.

\subsection{Dependence on age and metallicity}
\label{subsec:stellar_pop}

We again select stars on the bulge minor axis, $|l|<2\degrees$, within ${\Rgc<3.5\kpc}$. We use the same binning in $|b|$ used for the observations in the previous section, both for the observations and for the model. These are ${1.5\degrees<|b|<3.51\degrees}$, ${3.51\degrees<|b|<6.6\degrees}$ and ${7.13\degrees<|b|<8.85\degrees}$. We ignore the highest bin, centered on $|b|\sim12\degrees$, as it contains too few stars for a meaningful dissection by \feh.

We bin the selected stars by age and metallicity for the model and the APOGEE data respectively. For the model we use equal-step bins in age, from 4 to $10 \Gyr$, with a total of 40, 20 and 10 bins for the lowest, intermediate and highest latitude cuts, respectively. For the APOGEE data we use equal-number bins in metallicity, with 4 bins at the lowest and intermediate latitudes, and 3 at the highest.

\begin{figure}
    \centering
    \includegraphics[width=\columnwidth]{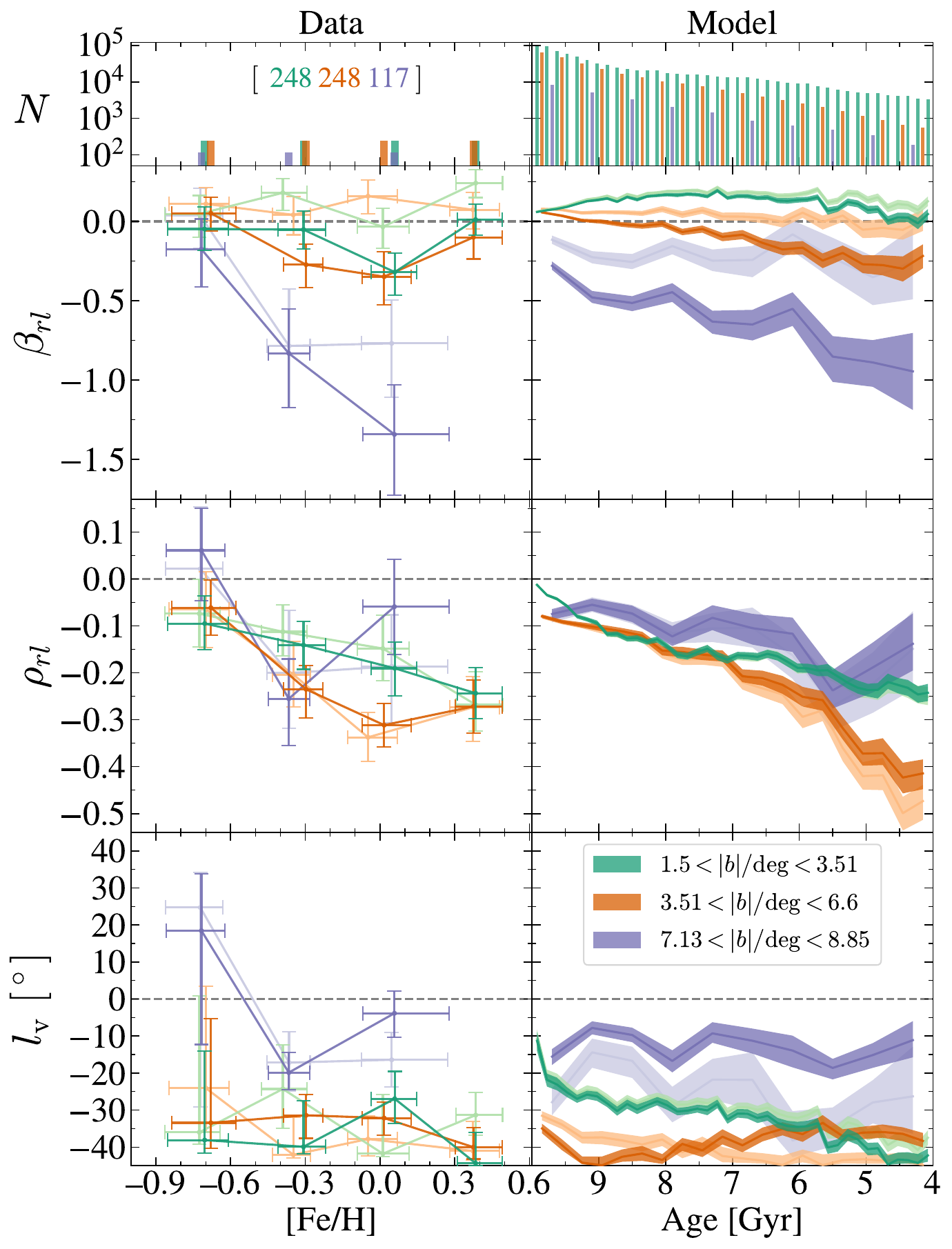}
    \caption{Kinematics as a function of metallicity (left) and age (right) for the APOGEE data and model respectively, along the bulge minor axis, $|l|<2\degrees$. Each panel shows 3 different latitudes, as indicated in the legend, with saturated and light colours representing results with $\Rgc<3.5\kpc$ and $\Rgc<2\kpc$, respectively. The number of stars in each bin is shown in the bar plots at the top.}
    \label{fig:kinematics_agemetal}
\end{figure}

Fig.~\ref{fig:kinematics_agemetal} shows the kinematics of the different metallicity (left) and age (right) bins for the APOGEE data and the model, respectively. We show the age increasing to the left in the panels representing the model, to ease the comparison (which is again purely qualitative). The APOGEE datapoints are shown at the median of each bin, with kinematic error bars measuring the $68\%$ bootstrap confidence interval, and metallicity errorbars indicating half the bin width in each direction. The model values are shown at the mid-point of each bin, and the $68\%$ bootstrap confidence intervals are joined into a shaded region.

From top to bottom we show $\ani$, $\corr$ and $\vertexabs$. The histogram at the very top shows the number of stars in each bin. The star numbers are also written as an inset for the observations.
In every panel we show in a lighter colour the results after reducing the radial cut to $\Rgc<2\kpc$. In the following descriptions we focus on the $\Rgc <3.5\kpc$ results (saturated colours).

The anisotropy of APOGEE stars and the model follow similar trends. It is negative and strong at the highest latitude (purple), increasing in magnitude towards more metal-rich and younger ages in the MW and in the model respectively. The values at the intermediate (orange) and lowest (green) latitudes are relatively isotropic compared to the highest latitude. At the intermediate latitude, the anisotropy of the model also exhibits a weak gradient with age, starting small and positive for the oldest stars, crossing $\ani = 0$ at ${\sim} 9\Gyr$, and decreasing towards younger ages. The anisotropy of the APOGEE stars at the intermediate latitude also grows slightly negative towards more metal-rich stars, although it then increases back to isotropy at the highest \feh. At the lowest latitude, $\ani$ is mostly flat across \feh\ and age. At this latitude, $\ani$ of the model is mostly positive, crossing the isotropic line at the youngest ages. Switching to ${\Rgc<2\kpc}$ leads to an increase in the anisotropy of most populations in the observations and the model, especially at higher latitudes. This is due to an increase in $\sigma_r$ and a decrease in $\sigma_l$. In the model $\sigma_r$ increases due to the cancellation of the positive and negative sides of the (fast) bar orbits, which dominate the velocity distribution at that reduced radial range (see Fig.~\ref{fig:xy_gal_vxvy}). As expected, the increase is more prominent for younger stars and at the latitudes where the X-shape is populated (orange and purple). The decrease in $\sigma_l$ in the model is due to the fact that $v_l$ increases away from $x=0$, as Fig.~\ref{fig:xy_gal_vxvy} shows.

Except for the metal-poor population in the APOGEE stars, the correlation (second row) is negative for all populations in the observations and in the model. The continuum of bar strengths seen in Fig.~\ref{fig:top_age_windows} manifests as a steady increase of the magnitude of the correlation with decreasing age, particularly at the lowest (green) and intermediate (orange) latitudes, with a larger gradient at the latter. The APOGEE data follow a similar trend, with stronger correlation at higher \feh\ at these latitudes. The correlation gradient with \feh\ is also larger at intermediate latitudes, although it seems to saturate at the highest \feh. At the highest latitude, the behaviour of the observations is more uncertain.
The effect of the radial change is relatively small on the $\corr$ of both the APOGEE data and the model. In this sense, $\corr$ is a robust probe of kinematic fractionation.

The vertex deviation (bottom row) in the model is negative and strong (${\lesssim-30\degrees}$) for most ages at the lowest (green) and intermediate (orange) latitudes. As explained in Section~\ref{sec:vd}, vertex deviation is very sensitive to non-zero correlations in isotropic regions. Therefore, even though the oldest stars are significantly less barred than the young ones, their anisotropy is small enough that their vertex deviation robustly reveals their non-axisymmetry. An exception are the stars older than ${\sim} 9 \Gyr$ at the lowest latitude, whose correlation in this hot region near the plane does drop to small enough values relative to the anisotropy that its vertex deviation eventually drops to $\sim -10\degrees$. The vertex deviation of the model at this lowest latitude is relatively unaffected by a change in the radial range to $\Rgc < 2\kpc$, resulting in just a slight decrease in magnitude for young stars. At the intermediate latitude, given the radial change affects where the anisotropy crosses the isotropy line, the vertex deviation curve shifts its lowest point from old to young stars. Regardless, the model's vertex deviation is strong and flat at this latitude for all ages. At the highest latitude, the anisotropy of the model is large enough that the vertex deviation is relatively weak, at $\vertexabs = -10\degrees$ to $-20\degrees$ for all populations. Reducing the radial range causes the anisotropy to drop in magnitude for all ages, causing the vertex deviation magnitude to increase to moderate values.

The vertex deviation of the APOGEE stars behaves similar to the model, with negative and relatively strong ($<-25\degrees$) values across all metallicities at the lowest (green) and intermediate (orange) latitudes, for both radial ranges, and more axisymmetric values at the highest (purple) latitude. At the lowest latitudes, even the most metal-poor stars have a very strong vertex deviation despite their small correlation, due to the small $\ani$. The large error is indicative of a nearly isotropic axisymmetry, especially at the intermediate and highest latitudes.

We have explored the kinematics of the model upon switching from Galactic to Cartesian cuts, and results were relatively unaffected, which means they are not dominated by projection effects. 

Additionally, we have recreated the results for the observational data in both Fig.~\ref{fig:kinematics_lat} and \ref{fig:kinematics_agemetal} upon applying a $5\%$ cap on the internal (see Section \ref{subsec:apogee-gaia}) fractional distance error, which discards ${\sim}19\%$ of the data, and the results remain largely unchanged, with all datapoints being consistent with the results shown here within a standard deviation.

Following on from the conclusion from the previous section, we find that the correlation is a great tracer of kinematic fractionation, \ie\ of varying bar strengths, at low and intermediate latitudes, for both the model and observations. This is in contrast with vertex deviation, which only serves as a blunt test for non-axisymmetry if populations are close to isotropic.


\section{Effect of distance uncertainties}
\label{sec:distance_error}

We explore the effect of distance errors in the precision and accuracy of the measurements of the velocity ellipses by introducing varying fractional uncertainties in the distance of stars in the model, using Monte Carlo error propagation with 500 repeats, assuming the distance uncertainties are Gaussian.

\begin{figure}
    \centering
    \includegraphics[width=\columnwidth]{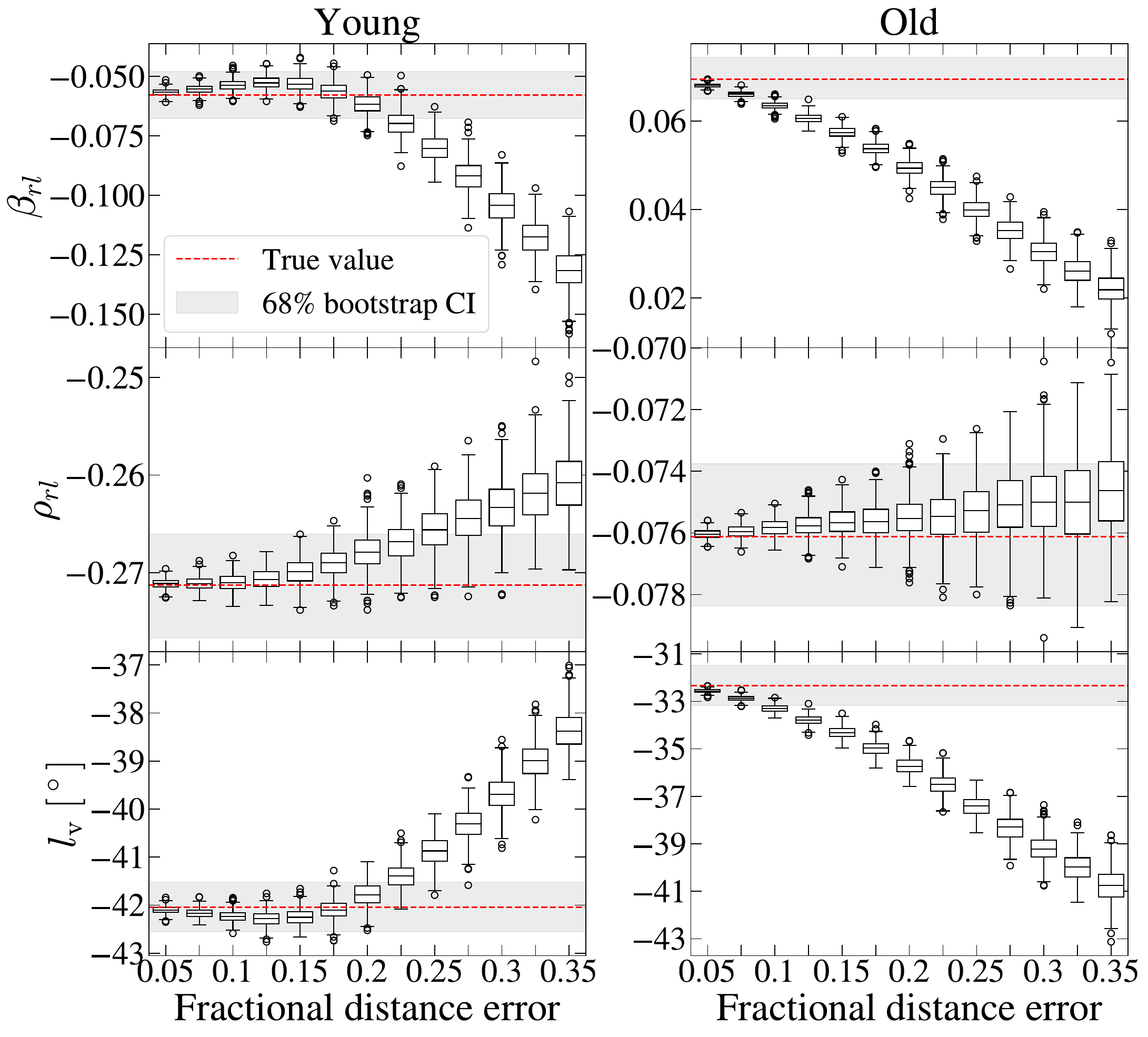}
    \caption{Monte Carlo distributions of anisotropy (top), correlation (middle) and vertex deviation (bottom) for the young (left) and old (right) populations along the bulge minor axis, resulting from increasing levels of fractional distance errors. Each distribution is represented by a boxplot, whose body spans the interquartile range (IQR), containing the central $50\%$ of the data, and is cut through by a solid line at the median. The boxplot whiskers span from ${\mathrm{Q}1-1.5\cdot \mathrm{IQR}}$ to ${\mathrm{Q}3+1.5\cdot \mathrm{IQR}}$, where $\mathrm{Q}1$ and $\mathrm{Q}3$ are the $25\%$ and $75\%$ percentile points. Any values beyond the whiskers are shown as individual markers. The red dashed lines show the true values of the statistics, with the $68\%$ bootstrap confidence interval shown as a grey shaded region.}
    \label{fig:distance_monte_carlo}
\end{figure}

Fig.~\ref{fig:distance_monte_carlo} shows the variations of the anisotropy (top), correlation (middle) and vertex deviation (bottom) with fractional distance errors from 0.05 to 0.35, for the young (left) and old (right) populations selected within ${|l|<2\degrees}$, ${3\degrees <|b|<6\degrees}$ and ${\Rgc<3.5\kpc}$. 

The variations of $\ani$ (and consequently $\vertexabs$) are dominated by relatively small changes in $\sig{l}$. After introducing the distance errors, the dependence of $v_l$ on $d$ leads to an increase in $\sig{l}$, and consequently a decrease in $\ani$. The anisotropy magnitude increases for the young population, which already has $\ani < 0$, while that of the old, which is initially positive, declines towards zero. This decrease begins immediately for the old population, but for the young once the $20\%$ error limit is reached. This is due to the differing radial density profiles of these populations. The density of the young population is less peaked at the GC than that of the old (see Fig.~\ref{subfig:top_density_youngold}), so young stars being lost\footnote{Stars originally at ${\Rgc>3.5\kpc}$ also make it inside the cut after the distance perturbation, but in smaller numbers as the density drops exponentially (see Fig.~\ref{subfig:top_density_youngold}).} at the edges of ${\Rgc<3.5\kpc}$ after the perturbation represent a more significant fraction of the population. Stars at the edges form part of the tails of the $v_l$ distribution. In addition, more stars with negative $v_l$ than positive are lost because the far and near sides contain mostly negative and positive $v_l$ stars respectively (see Fig.~\ref{fig:xy_gal_vxvy}) and further stars experience a stronger distance perturbation. The effect of this asymmetric loss is a decrease in dispersion which is significant enough to counteract the effect of the overall $\sig{l}$ broadening at the initial fractional errors for the young stars.
The correlation magnitude decreases with increasing distance uncertainties, although the relative change is smaller than that in the anisotropy. 
The increased anisotropy magnitude and weakened correlation leads to a reduction in the magnitude of the vertex deviation of the young stars past the $20\%$ error mark. Meanwhile, the anisotropy magnitude drop for the old stars causes its vertex deviation to grow in magnitude, reaching below $-40\degrees$, despite their very small correlation, which again illustrates the sensitivity of vertex deviation to non-zero correlations in isotropic populations.

\section{Velocity ellipses across the bulge}
\label{sec:lb_maps}

\begin{figure*}
    \begin{subfigure}{\textwidth}
        \centering
        \includegraphics[width=\textwidth]{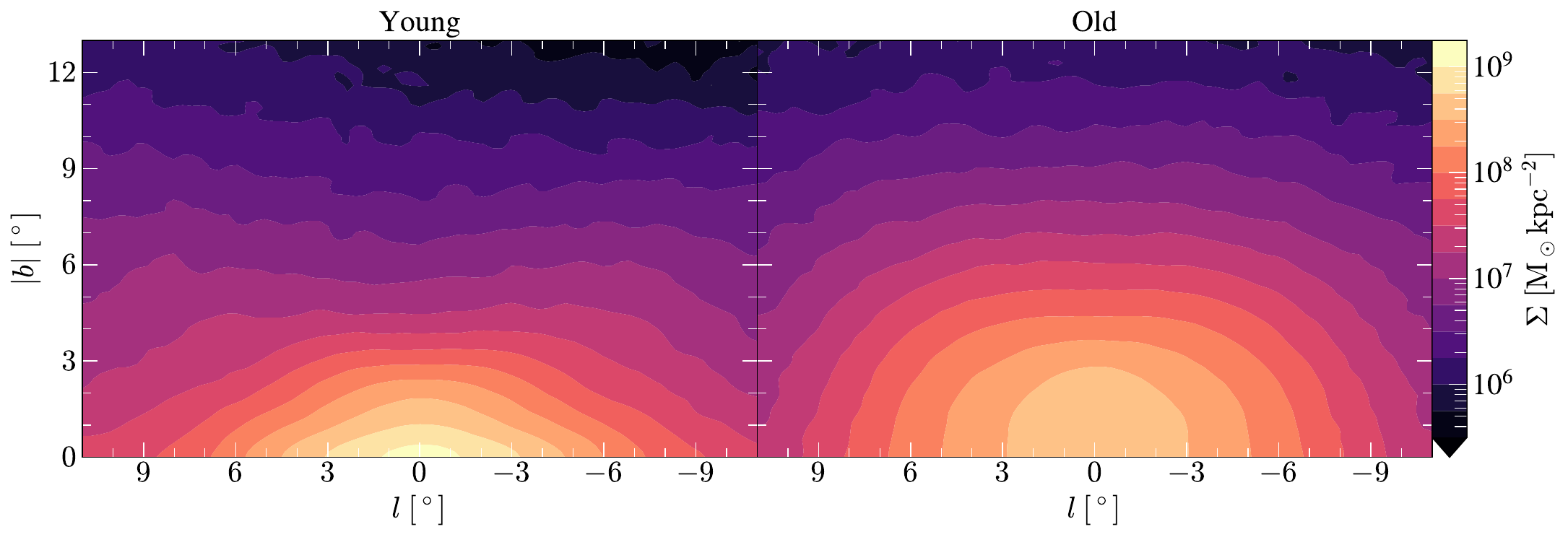}
        \caption{Model}
        \label{subfig:lb_model_density}
    \end{subfigure}
    \begin{subfigure}{\textwidth}
        \centering
        \includegraphics[width=\textwidth]{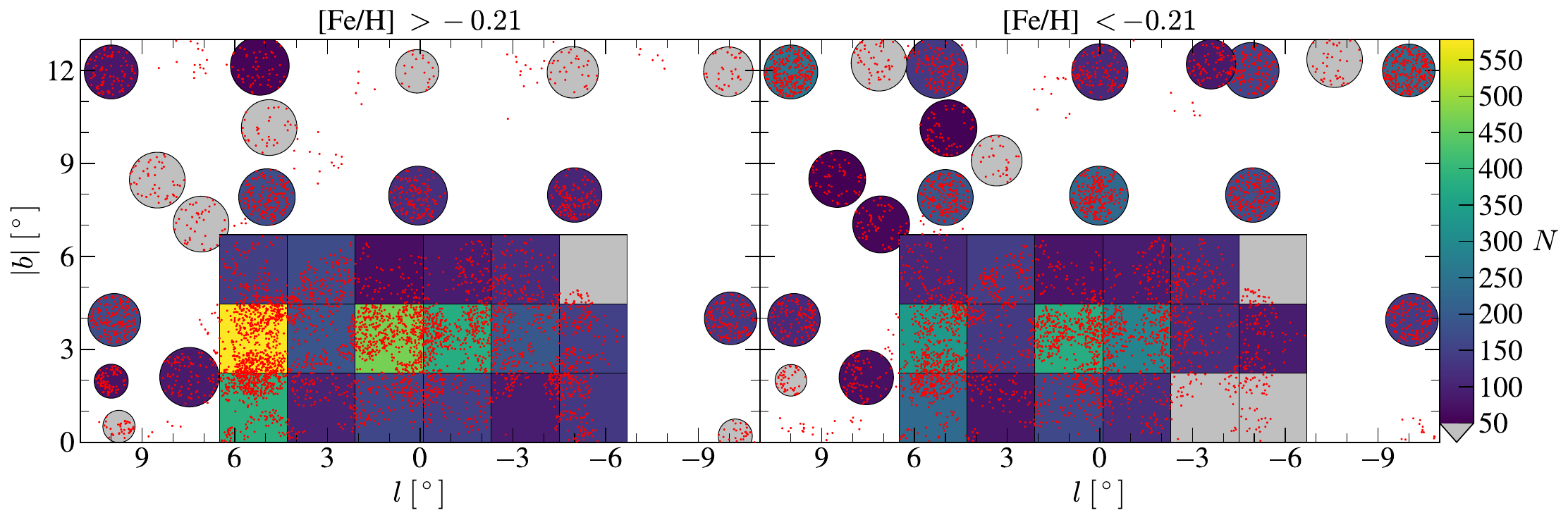}
        \caption{APOGEE data}
        \label{subfig:lb_obs_star_number}
    \end{subfigure}
    \caption{Spatial distributions of stars in $(l,b)$ space, within $\Rgc<3.5\kpc$. Panel (a) shows the surface density in the model for the young (left) and old (right) populations, and panel (b) shows the observed metal-rich (left) and metal-poor (right) star counts in selected bins. Grey bins in panel (b) did not make the $N\geq 50$ cut, so we disregard them in the kinematic maps of Section \ref{sec:lb_maps}. Individual observed stars are shown as red points.}
    \label{fig:lb_density}
\end{figure*}

In this section we consider the properties of the velocity ellipses for both the model and the APOGEE data across the entire bulge in Galactic coordinates, $(l,b)$. 
The use of Galactic velocities in fields away from $l=0\degrees$ introduces projection effects which obscure the interpretation of the vertex deviation values. Nonetheless, provided the line-of-sight distributions of observed stars is fairly representative of the actual ones, it is instructive to compare the model and the APOGEE data. Since the model is not a made-to-measure representation of the MW, we only compare the two qualitatively.

We consider stars within $\Rgc<3.5\kpc$. Fig.~\ref{subfig:lb_model_density} shows the surface density contours of the young (left) and old (right) stars in the model projected onto the $(l,b)$-plane. The young stars are very dense near the plane and develop an X-shape at ${|b|\gtrsim5\degrees}$. The arm of the X-shape at positive longitudes reaches slightly higher than the one at negative longitude, due to perspective. The old stars, instead, are thicker and boxy.

Given the heterogeneous density distribution of stars in the APOGEE data in the $(l,b)$-plane, as shown in the top left panel of Fig.~\ref{fig:apogee_windows}, we use continuous bins only in the densest region, within ${|l|,|b|\lesssim 6\degrees}$. For the rest of the stars we consider each pencil-beam field of view as a single bin. Fig.~\ref{subfig:lb_obs_star_number} shows the resulting star number in each bin, with individual stars in red. In our analysis we consider only bins containing at least 50 stars. 

Given that we are considering the full range of latitudes ${|b|<13\degrees}$, down to the Galactic plane, the nuclear disc might affect our results closest to the plane. In the model we reduce its contribution using our age cuts (the model's nuclear disc is younger than $4\Gyr$). Its effect is also small in the observations given most of our stars in our final bins are located at $|b|>0.5\degrees$ (see top left of Fig.~\ref{fig:apogee_windows}), at which point the contamination from it is no larger than $25\%$ according to the modelling of \citet{sormani2022}.

\begin{figure*}
    \begin{subfigure}{\textwidth}
        \centering
        \includegraphics[width=\textwidth]{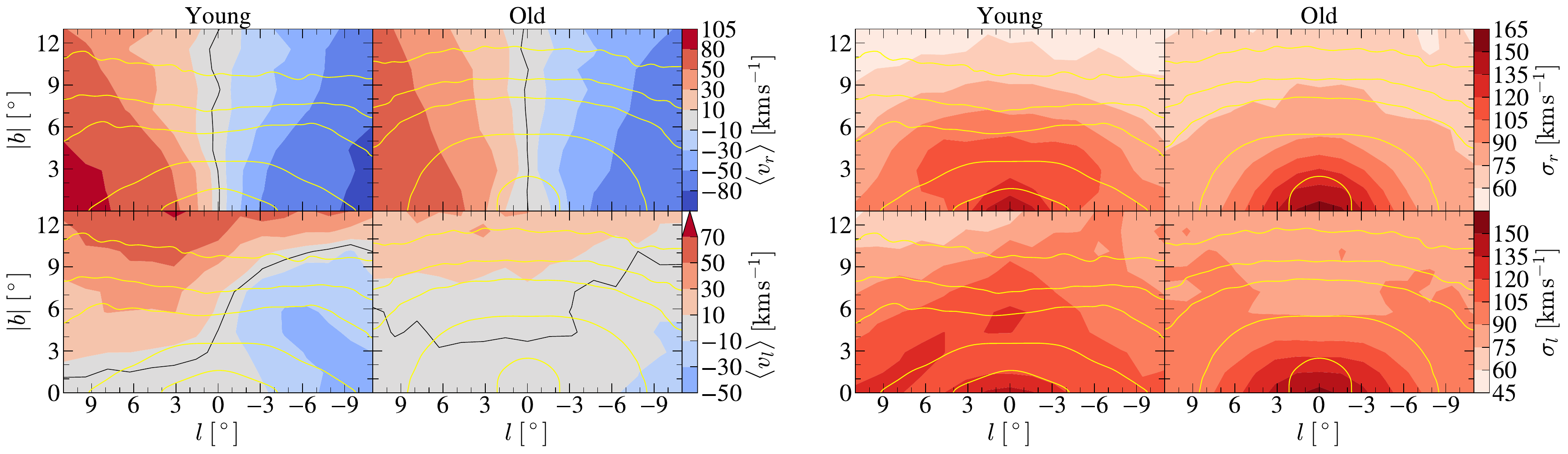}
        \caption{Model}
        \label{subfig:lb_model_vxvy}
    \end{subfigure}
    \begin{subfigure}{\textwidth}
        \centering
        \includegraphics[width=\textwidth]{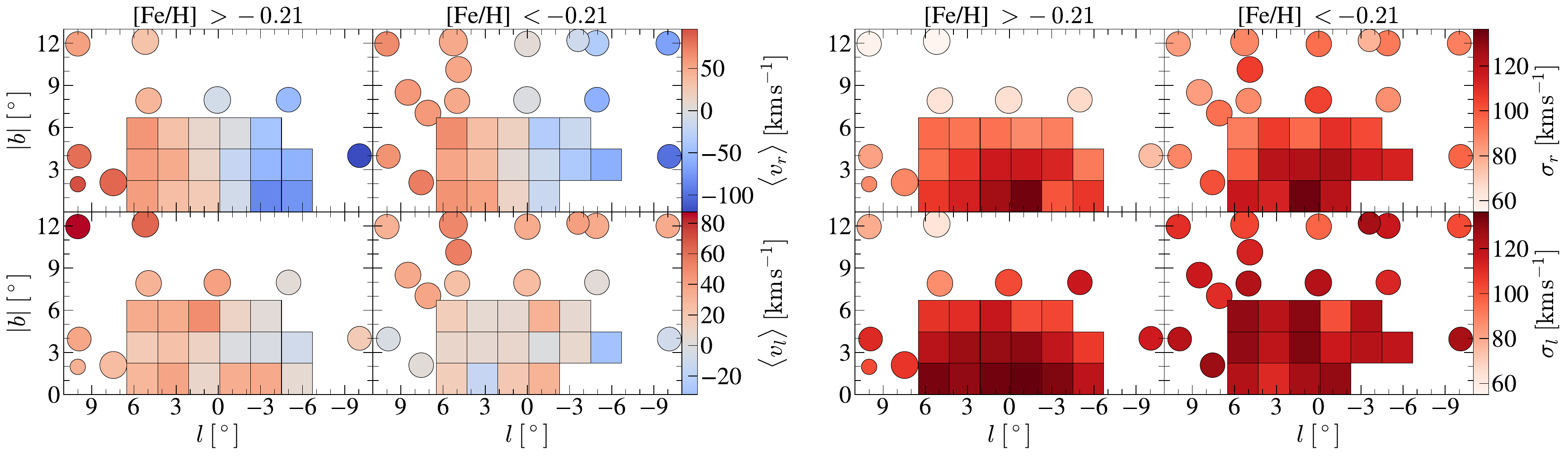}
        \caption{APOGEE data}
        \label{subfig:lb_obs_vxvy}
    \end{subfigure}
    \caption{Kinematic maps in Galactic coordinates of the Galactic velocities for (a) the model (see Fig.~\ref{fig:xy_gal_vxvy} for the face-on view) and (b) the APOGEE data, showing the mean velocities (left) and the dispersions (right). In the maps for the model the black contours follow values of zero, while the yellow contours outline the density distribution. In each block, the top and bottom rows show radial and tangential motions, while the left and right columns correspond to young/metal-rich and old/metal-poor stars.}
    \label{fig:lb_vxvy}
\end{figure*}

Fig.~\ref{fig:lb_vxvy} shows the mean Galactic velocities and their dispersions. For both (a) model and (b) observations, $\meanvr$ shows the projection of the clockwise rotation, with a gradient towards decreasing $|b|$ and increasing $|l|$. The contours of the young stars are more pinched than those of the old, exhibiting faster rotation at lower latitudes. The near and far sides of the young stars' forbidden velocities (see Fig.~\ref{fig:xy_gal_vxvy}) have mostly cancelled out. In the $\meanvl$ panels, the rotation of the near and far sides have largely cancelled out, but an asymmetric projection effect remains in both the model and the APOGEE data. This effect is weaker for the old and metal-poor stars than for the young and metal-rich stars, due to projection effects of the near and far ends of the bar.

\begin{figure*}
    \begin{subfigure}{\textwidth}
        \centering
        \includegraphics[width=\textwidth]{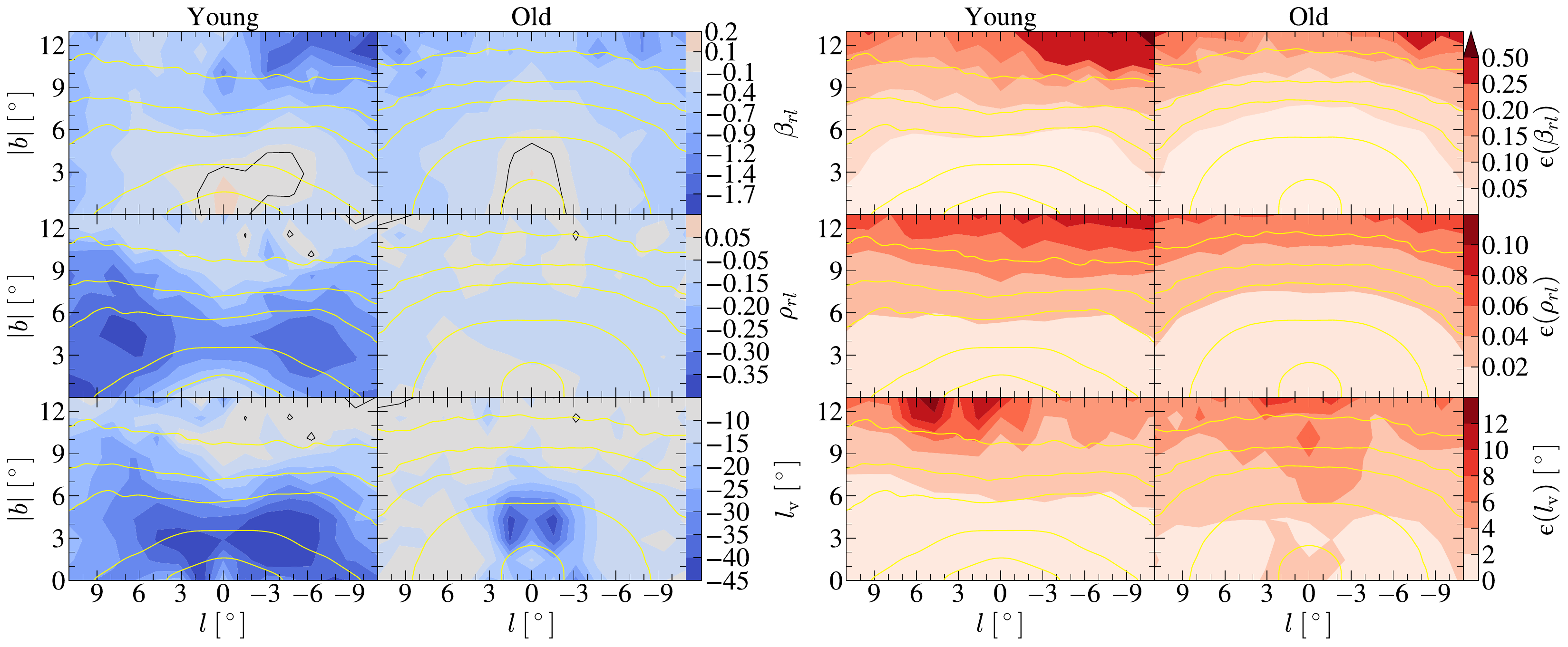}
        \caption{Model}
        \label{subfig:lb_model_anicorr}
    \end{subfigure}
    \begin{subfigure}{\textwidth}
        \centering
        \includegraphics[width=\textwidth]{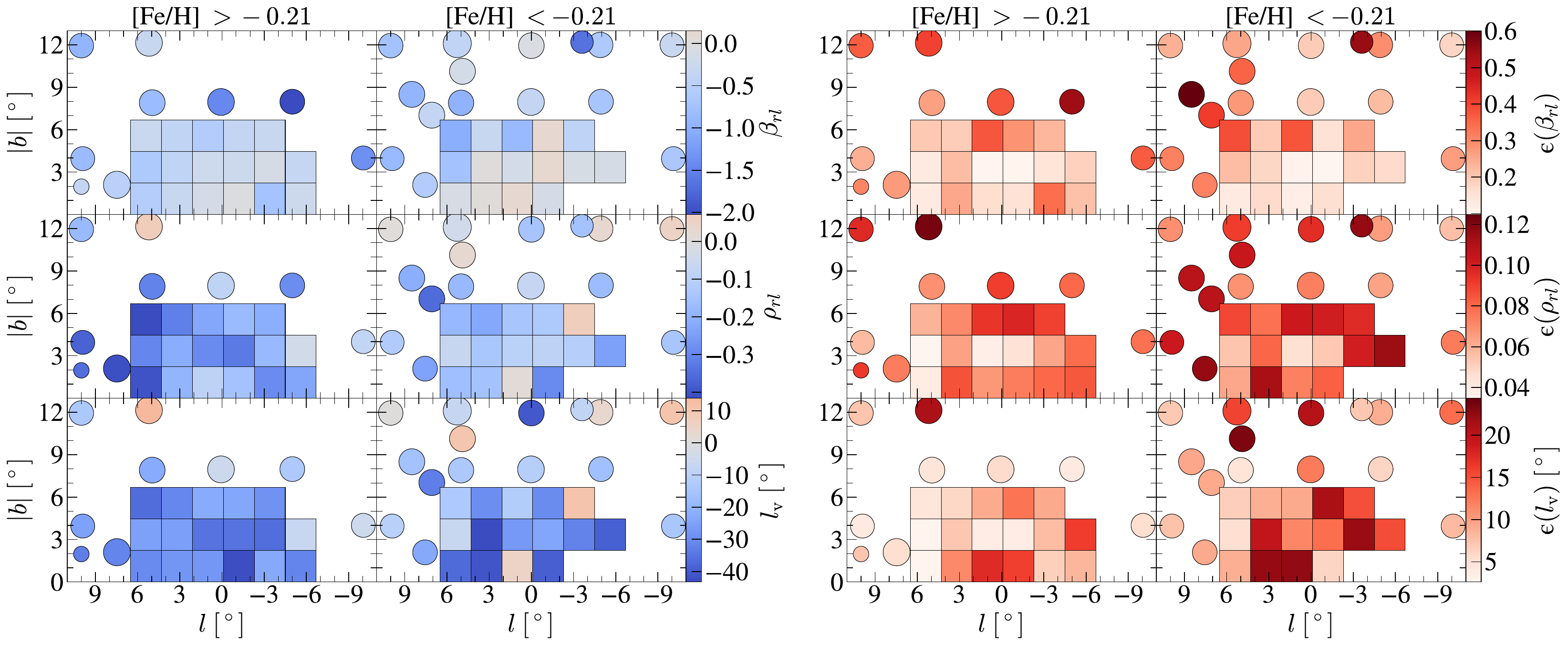}
        \caption{APOGEE data}
        \label{subfig:lb_obs_anicorr}
    \end{subfigure}
    \caption{Same as Fig.~\ref{fig:lb_vxvy} but now showing the anisotropy, correlation, and vertex deviation.}
    \label{fig:lb_anicorr}
\end{figure*}

Both the model and the APOGEE data generally show $\sigma_l>\sigma_r$, and both dispersions grow with decreasing latitudes. The face-on view of the model in Fig.~\ref{fig:xy_gal_vxvy} shows that at lines of sight within $|l|\lesssim5\degrees$, $\sigma_r$ is larger than $\sigma_l$, especially for the young stars, whereas outside $\sigma_l$ tends to be larger. However, the effect of near- and far-side $v_l$ values cancelling out has led to a larger $\sigma_l$ than $\sigma_r$ everywhere except in the very inner region in Fig.~\ref{subfig:lb_model_vxvy}. This is reflected in the isotropic regions near the galactic centre in the first row of Fig.~\ref{subfig:lb_model_anicorr}. Both young and old stars show $\ani>0$ under the isotropic region, as can be seen also in Fig.~\ref{fig:kinematics_agemetal} along the minor axis (green line), and $\ani<0$ everywhere else in the $(l,b)$-plane. There are hints of this behaviour in the APOGEE data, shown in the first row of Fig.~\ref{subfig:lb_obs_anicorr}, which are isotropic in the central region near the plane, and have $\ani < 0$ everywhere else.

Fig.~\ref{fig:xy_gal_anicorr} showed that the $\corr<0$ and $\vertexabs<0\degrees$ poles of the quadrupoles of the young stars in the model viewed in the $(x,y)$-plane dominate all lines of sight, and are particularly strong at the X-shape overdensities. The second and third rows of Fig.~\ref{subfig:lb_model_anicorr} show how this translates to the $(l,b)$-plane. We have $\corr<0$ and $\vertexabs<0\degrees$ mostly everywhere. The correlation is particularly strong at intermediate latitudes, and at longitudes where the X-shape resides $(|l|\gtrsim5\degrees)$, particularly on the near side ($l>0\degrees$) of the bar, which is also where 
\citet{gough-kelly2022} found that the forbidden velocities dominated. We find an agreement with the correlation of the metal-rich population in the observations (second row of Fig.~\ref{subfig:lb_obs_anicorr}). The vertex deviation of the young population in the model is particularly strong in the inner isotropic regions, particularly at $l<0\degrees$ where the isotropic region is more extended. The vertex deviation of the metal-rich stars (third row of Fig.~\ref{subfig:lb_obs_anicorr}) follows a similar trend, with strong negative values in the inner region.

The correlations of the old stars in the model and of the metal-poor ones in the APOGEE data are also negative mostly everywhere, but the values are smaller compared to the young and metal-rich stars. Despite this, given the isotropy in the inner region, the vertex deviation of the old population in the model is strong at $|b|\sim4\degrees$ on either side of the bulge minor axis. The vertex deviation of the metal-poor stars in APOGEE also shows some strong values in the inner region. 
However, some bins also show high vertex deviation errors, indicative of isotropic axisymmetry. The metal-poor population has fewer stars in the ${|b|<6\degrees}$ region than the metal-rich (see Fig.~\ref{subfig:lb_obs_star_number}); more stars with future surveys would help ascertain the behaviour there.


\section{Discussion}
\label{sec:discussion}

We have used the high resolution $N$-body + SPH simulation from \citet{Debattista2017} to study age-dependent kinematic trends across the bulge. In this model, all stars form out of cooling gas, which results in correlations between the ages of stars and their kinematics. Once the bar forms, the stellar populations separate through kinematic fractionation, which leads to the young populations being strongly barred and X-shaped, while the older populations are weakly barred and boxy. The model, which has evolved purely in isolation, is able to explain trends observed in the velocity ellipses of the MW's bulge.

\subsection{Dependence of bar strength on [Fe/H] in the MW}

Fig.~\ref{fig:kinematics_agemetal} shows the variation of the correlation, \corr, with \feh\ in the APOGEE DR16 $+$ \gaia\ DR3 data. At $3\degrees < |b| < 6\degrees$, which our model suggests is the best location for tracing the bar's amplitude, the overall trend appears to be a rise in $|\corr|$ (becoming increasingly negative) from zero at $\feh \sim -0.7 \dex$ until $\feh \sim 0$. This is the first indication that the strength of the bar in the MW varies smoothly with \feh, as predicted by the model due to kinematic fractionation. 
At the highest \feh\ bin, \corr\ appears to plateau or even decline in amplitude slightly. However, the number of such metal-rich stars is low so the uncertainty is larger than the depth of the minimum. Better observations will help establish whether the growing trend extends to the most \feh-rich stars.

\subsection{Future prospects}

Future observations, particularly with next-generation surveys like MOONS \citep{gonzalez2020}, 4MOST \citep{bensby2017}, and VRT-LSST, will provide a much more detailed and comprehensive picture of the Galactic bulge. The high multiplexing capabilities of MOONS \citep{cirasuolo2020} and 4MOST will enable large-scale spectroscopic surveys of the bulge, providing valuable kinematic data for different stellar populations that can be compared with our model predictions. In particular, a Galactic Guaranteed Time Observations survey \citep{gonzalez2020} will devote 100 nights to map the central regions of the MW and some of its satellites. The $|l| \leq 10\degrees$ and $|b|\leq 4\degrees$ region (in addition to the adjacent disc at negative longitudes) will be sampled with a very dense and contiguous grid of fields (about 1000 targets across 25 arcmin FoV) obtaining spectra in the RI, YJ and H bands. The observations in the high resolution H band will pierce through the dust in the Galactic mid-plane and sample this key region with an unprecedented sample size of about 0.5 million red clump stars. This data will be complemented with NIR photometry and proper motions from the VVV survey \citep{VVV}. Although the precision of the proper motions will not be as high as future {\it Gaia} data releases, they will be available for these highly reddened regions where {\it Gaia} observations are shallower. These datasets combined will provide the unique opportunity to examine in detail, and with statistically significant samples, the kinematic signatures described here. They will be instrumental to settle the debate concerning the origin and evolution of the bulge stellar populations. In addition, VRT will provide photometric data for billions of stars, helping to identify target stars for follow-up spectroscopic observations and enabling detailed structural mapping of the bulge. This will allow for a more comprehensive distribution of observational data across the bulge. Such data will be invaluable for refining our understanding of the kinematics of the Galactic bulge. 

Our results provide guidance for future work using velocity ellipses at the bulge. The velocity ellipses can be parameterised by the dimensionless in-plane anisotropy, $\ani$, correlation, $\corr$, and vertex deviation, $\vertexabs$ (though these are not independent, see Appendix~\ref{appendix:anicorr}). Non-zero vertex deviation has been used as a tracer of non-axisymmetric structure; however, when $\ani \sim 0$, \ie\ when $\sig{l} \sim \sig{r}$, then $\vertexabs$ generally takes values near the maximum of $\pm 45\degrees$. The region around $|b| \sim 3\degrees$ to $6\degrees$ (see Fig.~\ref{fig:kinematics_lat} and \ref{fig:kinematics_agemetal}) in both the model and APOGEE suffers from this effect when all stars are considered. The relative constancy of $\vertexabs$ across \feh, only dropping suddenly to $\vertexabs \sim 0\degrees$ at $\feh \sim -0.7 \dex$ \citep{Soto_2007}, has in the past been interpreted as evidence that a metal-poor unbarred, and probably accreted, population is present, alongside a population with a relatively uniformly strong bar. However, the lack of variation in the model's $\vertexabs$ as a function of age, seen in Fig.~\ref{fig:kinematics_agemetal}, does not reflect a constancy in bar strength: \citet{Debattista2017} showed that the bar strength varies with age in the same model as we use here. Likewise, \citet{Debattista2019} using a cosmological simulation from the FIRE-2 suite \citep{hopkins+2018}, showed that even though the bar strength changed quite substantially for stars between $1\Gyr$ to $9\Gyr$ old, $\vertexabs$ was constant throughout this age range. Additionally, in Appendix~\ref{appendix:weaker_bars} we compare the model presented in this paper with two additional models, containing a weaker bar and an oval respectively, and find an equally large $\vertexabs$ peak amplitude for the young populations in the strong and weaker bar models. We conclude that $\vertexabs$ is actually a poor tracer of bar strength. 

On the other hand, $\ani$ is very sensitive to the distance range employed, and is in general weakly varying with age and \feh\ for stars at $|b| \lesssim 4\degrees$. Instead, the correlation, $\corr$, changes monotonically with age and bar strength. The variation of $\corr$ in the APOGEE data increases monotonically in amplitude with \feh\ at $1.5\degrees < |b| < 3.5\degrees$, but at $3.5\degrees < |b| < 6.6\degrees$ it may flatten at the highest \feh, though it is still consistent with increasing amplitude within the errors. 
Moreover, $\corr$ is only very weakly dependent on the distance range used (which is evident in the model and confirmed by the APOGEE data), which makes it robust to distance uncertainties in observational data. We have also verified that selecting stars based on their heliocentric distance, such as $6.1 < d/\kpc < 10.1$, still changes $\corr$ mildly. This is also true if we use distance cuts which are asymmetric (\eg\ $7.1 < d/\kpc < 10.1$ and $6.1 < d/\kpc < 9.1$). Thus observational selection functions that are metallicity dependent do not severely affect the comparison of \corr\ for different populations. 

In addition, in Appendix~\ref{appendix:bootstrap_assumption} we test the main assumption of the bootstrap and show that it breaks for vertex deviation for sample sizes smaller than $n\lesssim 300$, and that \vertexabs\ exhibits biases (which is the average difference between the value of a population and of samples extracted from it) as large as $10\degrees$-$20\degrees$. On the contrary, the correlation is well-behaved, with a standard error decreasing as $n^{-2}$ and no bias even at sample sizes as small as $n=50$.

Thus we propose that future studies should devote more attention to measuring the correlation, $\corr$, to explore whether the MW's bar amplitude is a continuous function of \feh. The region around $|b| \sim 3\degrees-6\degrees$ along the bulge's minor axis is particularly promising as a location for exploring the variations with \feh. At larger heights, the mixing between Galactocentric radial and vertical motions reduces the strength of the correlation and are therefore not ideal for probing the variation of the bar's strength.

\subsection{Summary}

We have computed the kinematics of a young, strongly barred and peanut-shaped population, and an old, weakly barred and boxy population in a model of a barred galaxy evolving purely secularly. 
The old stars have nearly axisymmetric velocities while large bar streaming motions are present in the young population, as previously shown by \citet{gough-kelly2022}. 
Here we have focused in particular on the in-plane anisotropy, correlation and vertex deviation of the velocity ellipses of the different populations. We have compared these results with data from APOGEE DR16 cross-matched with {\it Gaia} DR3. Our main results are as follows:
\begin{enumerate}
\item The dispersions of the galactocentric cylindrical velocities in the model exhibit quadrupoles in the $(x,y)$ plane. The \sig{R}\ quadrupole is elongated along the bar direction while that in \sig{\phi}\ is perpendicular to it. The overall velocity ellipse has a quadrupole in the anisotropy, \anicyl, which is also aligned with the bar, while the quadrupoles in the correlation, \corrcyl, and the vertex deviation, \vertexabsRphi, are rotated by $45\degrees$ ahead of the bar. All of these quadrupoles are stronger in the young population than in the old one. (See Fig.~\ref{fig:xy_cyl_vxvy} and \ref{fig:xy_cyl_anicorr}.)
\item Switching to heliocentric velocities, $v_r$ and $v_l$, the $(x,y)$ maps of the model's resulting dispersions, $\sigma_r$ and $\sigma_l$, appear as perspective-distorted versions of the intrinsic $\sigma_R$ and $\sigma_\phi$ maps, with comparable peaks and quadrupoles.
While the $(x,y)$-maps of anisotropy, correlation, and vertex deviations retain strong quadrupoles in heliocentric velocities, their orientations are quite different to those of the galactocentric cylindrical velocities. The quadrupoles in the anisotropy and in the correlation in the old population are weaker than in the young one, whereas those in the vertex deviation are comparable. The positive poles of the anisotropy quadrupole and the negative poles of the correlation and vertex deviation quadrupoles are bridged together for the young stars, while the bridge is much shallower in the old stars. This bridging is what gives rise to the negative correlation and vertex deviation when one looks along $l=0\degrees$. (See Fig.~\ref{fig:xy_gal_vxvy} and \ref{fig:xy_gal_anicorr}.)
\item Along the minor axis ($|l| \lesssim 2\degrees$), the APOGEE data and the model follow similar trends with $|b|$ in \ani, \corr, and \vertexabs. In particular, we find isotropic \ani\ values below $|b| \lesssim 5\degrees$, and $\ani < 0$ beyond that both in the \feh-rich/young strongly-barred population, which has the larger \ani\ amplitudes, and in the \feh-poor/old weakly-barred one. The \corr\ amplitudes are larger in the \feh-rich/young population than in the \feh-poor/old one, with a negative U-shaped vertical profile in both. In spite of these differences, \vertexabs\ reaches a strong peak amplitude, near $-45\degrees$, within $|b|<6\degrees$ for the two populations. This highlights that \vertexabs\ is a blunt probe of non-axisymmetry when populations are isotropic, unlike \corr\ which does vary with bar strength. Moreover, changing the distance selection function changes \ani\ and \vertexabs, but has only a weak effect on \corr. (See Fig.~\ref{fig:kinematics_lat}.)
\item Along the minor axis, at fixed $|b|$, \ani\ and \corr\ vary with \feh\ (APOGEE data) and age (model) but \vertexabs\ is relatively constant. The increasing amplitude of \corr\ with \feh\ in the APOGEE data suggest that in the MW the bar amplitude varies continuously, as predicted by kinematic fractionation, rather than being constant above some metallicity. (See Fig.~\ref{fig:kinematics_agemetal}.)
\item We simulated line-of-sight distance uncertainties in the model along the bulge minor axis and found that, in the young stars, the effect is negligible in the anisotropy, correlation and vertex deviation for fractional errors less than $20\%$. For the old stars, the amplitude of the anisotropy decreases and, as a consequence, the vertex deviation reaches values as strong as in the young population. The correlation is the least affected in both populations, barely deviating from the original value, even with distance uncertainties as large as $35\%$. (See Fig.~\ref{fig:distance_monte_carlo}.)
\item Projected on to the full $(l,b)$-plane, the correlation is mostly negative everywhere for the young and old stars. However, while it is uniformly weak for the old stars, the correlation of the young is stronger, particularly on the near side of the bar ($l>0\degrees$), and exhibits two strong peaks near the arms of the X-shape.
The vertex deviation is strong and negative for the young population, with two peaks closer to the minor axis compared to those of the correlation. The old stars have weak vertex deviation except very near the minor axis, where $\ani\sim0$, exhibiting a small-scale but strong double-peak. The maps of the APOGEE data for the \feh-rich and \feh-poor stars share some of these properties, including a larger amplitude \corr\ in \feh-rich stars, with stronger values at $l>0\degrees$, and relatively isotropic distributions in both populations near the Galactic Centre. However more data are required to confirm these trends. (See Fig.~\ref{fig:lb_anicorr}.)
\item We have highlighted the promise that measurements of the correlation between $v_r$ and $v_l$ along the bulge minor axis hold. The amplitude of the correlation of a given population increases with bar strength, is robust to variations in the distances probed, and is detectably varying in the MW. We have shown that the vertex deviation is instead a poor tracer of bar strength. 
\end{enumerate}


\section*{Acknowledgements}

S.~G.~K.~ acknowledges support from the Moses Holden Studentship, with particular thanks to Patrick Holden. We thank the referee, Iulia Simion, for comments that helped improve the paper. The simulation used in this paper was run at the High Performance Computing Facility of the University of Central Lancashire. A.~R.~A. acknowledges support from DICYT through grant 062319RA and from ANID through FONDECYT Regular grant No. 1230731. L.~B. e S. acknowledges the support of the Heising Simons Foundation through the Barbara Pichardo Future Faculty Fellowship from grant \# 2022-3927.

This work presents results from the European Space Agency (ESA) space mission {\it Gaia}. {\it Gaia} data are being processed by the Gaia Data Processing and Analysis Consortium (DPAC). Funding for the DPAC is provided by national institutions, in particular the institutions participating in the Gaia MultiLateral Agreement (MLA). The Gaia mission website is \url{https://www.cosmos.esa.int/gaia}. The {\it Gaia} archive website is \url{https://archives.esac.esa.int/gaia}.

Funding for the Sloan Digital Sky Survey IV has been provided by the Alfred P. Sloan Foundation, the U.S. Department of Energy Office of Science, and the Participating Institutions. SDSS acknowledges support and resources from the Center for High-Performance Computing at the University of Utah. The SDSS web site is \url{www.sdss4.org}.

SDSS is managed by the Astrophysical Research Consortium for the Participating Institutions of the SDSS Collaboration including the Brazilian Participation Group, the Carnegie Institution for Science, Carnegie Mellon University, Center for Astrophysics | Harvard \& Smithsonian (CfA), the Chilean Participation Group, the French Participation Group, Instituto de Astrofísica de Canarias, The Johns Hopkins University, Kavli Institute for the Physics and Mathematics of the Universe (IPMU) / University of Tokyo, the Korean Participation Group, Lawrence Berkeley National Laboratory, Leibniz Institut f\"ur Astrophysik Potsdam (AIP), Max-Planck-Institut f\"ur Astronomie (MPIA Heidelberg), Max-Planck-Institut f\"ur Astrophysik (MPA Garching), Max-Planck-Institut f\"ur Extraterrestrische Physik (MPE), National Astronomical Observatories of China, New Mexico State University, New York University, University of Notre Dame, Observat\'orio Nacional / MCTI, The Ohio State University, Pennsylvania State University, Shanghai Astronomical Observatory, United Kingdom Participation Group, Universidad Nacional Aut\'onoma de M\'exico, University of Arizona, University of Colorado Boulder, University of Oxford, University of Portsmouth, University of Utah, University of Virginia, University of Washington, University of Wisconsin, Vanderbilt University, and Yale University.


\section*{Data availability}

The simulation data underlying this article will be shared on reasonable request. The APOGEE DR16 data is publicly available at \url{https://live-sdss4org-dr16.pantheonsite.io/irspec} and {\it Gaia} DR3 data are publicly available at \url{https://www.cosmos.esa.int/web/gaia/data-release-3}.




\bibliographystyle{mnras}
\bibliography{allrefs} 



%
\appendix

\section{Vertex deviation}
\label{vd_appendix}

Vertex deviation is computed in two different ways in the literature, causing some confusion in its interpretation. For the benefit of future studies, in this appendix we review the two definitions and show the difference between them.

\subsection{Derivation}

The velocity dispersion tensor, Eqn.~\ref{eq:dispersion_tensor}, in 2D can be expressed as the covariance matrix (see appendix in \citealt{smith_2009}), $\mathcal{C}$, which contains the $i^{\rm th}$ and $j^{\rm th}$ variances along the diagonal, and the covariance in the off-diagonal terms. Given one of the velocity ellipse's principal axes, $\vbold$, with dispersion $\sigma$ along its direction, the eigenvalue problem can be set up as $\mathcal{C}\vbold = \sigma\vbold$, or
\begin{equation}
    \begin{pmatrix}
    \vari & \covij \\
    \covij & \varj
    \end{pmatrix}
    \begin{pmatrix}
    v_i \\ v_j
    \end{pmatrix} = \sigma \begin{pmatrix}
    v_i \\ v_j
    \end{pmatrix},
\end{equation}
which is a system of equations that can be combined to yield
\begin{equation}
    \frac{\varj - \vari}{\covij} = \frac{v_j}{v_i} - \frac{v_i}{v_j}.
    \label{eq:eigenvector_components}
\end{equation}
Calling $\vertexabs$ the angle that $\vbold$ makes with axis $i$, and $\xi$ its complementary angle, we can re-write the equality above as
\begin{align}
    \frac{\varj-\vari}{\covij}\nonumber  &= \tan{\vertexabs}-\tan{\xi} \\
                                &= (1+\tan{\vertexabs}\tan{\xi})\tan(\vertexabs-\xi) \nonumber \\
                                &= 2\tan(2\vertexabs-\pi/2) \label{eq:tandiff} \\
                                &=-2\cot(2\vertexabs), \label{eq:cot}
\end{align}
having used the trigonometric identity for the tangent of a difference. Eqn.~\ref{eq:tandiff} follows from writing $\tan \vertexabs \tan \xi = 1$ and $\xi = \pi/2 - \vertexabs$\footnote{To be precise, if $\vertexabs<0\degrees$ we should write ${\xi = -(\pi/2 + \vertexabs)}$. This would give ${\vertexabs-\xi=2\vertexabs+\pi/2}$, changing Eqn.~\ref{eq:tandiff}. However, given ${\tan(x)=\tan(x\pm\pi)}$, Eqn.~\ref{eq:cot} would still hold.}. Lastly, we used the trigonometric relations between complementary angles to write Eqn.~\ref{eq:cot}, which we can recast as
\begin{equation}
    \tan(2\vertexabs) = \frac{2\covij}{\vari -\varj}.
    \label{eq:standard}
\end{equation}

Our derivation is an alternative to the one provided by \citeauthor{BM98} (\citeyear{BM98}, page 630), which is the one typically referenced, in which they reach the same result but along the way (Eqn.~10.15 to Eqn.~10.16) introduce a negative sign, which we did not need here.

\subsection{Vertex deviation definitions}

In Eqn.~\ref{eq:eigenvector_components} we absorbed the eigenvalue $\sigma$, which means Eqn.~\ref{eq:standard} holds for both eigenvectors of $\mathcal{C}$. 

It is in the calculation and interpretation of $\vertexabs$ from Eqn.~\ref{eq:standard} that authors differ. One way to compute it, as used by several studies \citep[\eg][]{zhao1994,dehnen&binney1998,smith_2009,roca-fabrega2014,Budenbender2015} is
\begin{equation}
    \vertex = \frac{1}{2}\mathrm{atan2}\left(2\covij,\vari-\varj\right),
    \label{eq:atan2}
\end{equation}
where 
\begin{equation}
    \mathrm{atan2}(y,x) = \begin{cases}
     \arctan(y/x), & \text{if}\ x\geq0\\
     \arctan(y/x)+\mathrm{sign}(y)\pi & \text{if}\ x<0
    \end{cases}.
    \label{eq:atan2_def}
\end{equation}
Eqn.~\ref{eq:atan2_def} is equivalent to writing $\mathrm{atan2}(y,x) \equiv \mathrm{Arg}(x+iy)$, where $\mathrm{Arg}$ is the principal argument of the complex number $x+iy$. Selecting the principal branch gives values in the range $(-180,180]\degrees$. Therefore Eqn.~\ref{eq:atan2} limits $\vertex$ to the range $(-90,90]\degrees$, and measures the {\it angle that the semi-major axis of the velocity ellipse makes with the positive side of the horizontal axis direction}, $\ihat$.

We have used a tilde in Eqn.~\ref{eq:atan2} to differentiate it from the other definition \citep[\eg][]{Soto_2007,soto2012,Babusiaux2010,Debattista2019,simion2021}, which we have used throughout this paper. It is given by Eqn.~\ref{eq:standard_abs}, which involves taking the absolute value of the difference between the dispersions. Taking the absolute value removes the ambiguity in the calculation of $\vertexabs$ from Eqn.~\ref{eq:standard}, allowing us to apply the standard $\arctan$ function. The resulting values live in the range $[-45,45]\degrees$, and measure the {\it angle that the semi-major axis of the velocity ellipse makes with the coordinate axis it is closest to}, with the same sign as the correlation (and as $\vertex$). 

We can convert between the two definitions using
\begin{equation}
    \vertexabs = \begin{cases}
     \vertex, & \text{if}\ \vari\geq\varj\\
     \mathrm{sign}(\vertex)(90\degrees-|\vertex|) & \text{if}\ \vari<\varj
    \end{cases},
    \label{eq:vertex_to_abs}
\end{equation}
and vice-versa. In words, they are the same except when the anisotropy (Eqn.~\ref{eq:anisotropy}) is negative, in which case $|\vertex|>45\degrees$ and $\vertexabs$ is its complementary angle, with the same sign. 

\begin{figure}
    \centering
    \includegraphics[width=\columnwidth]{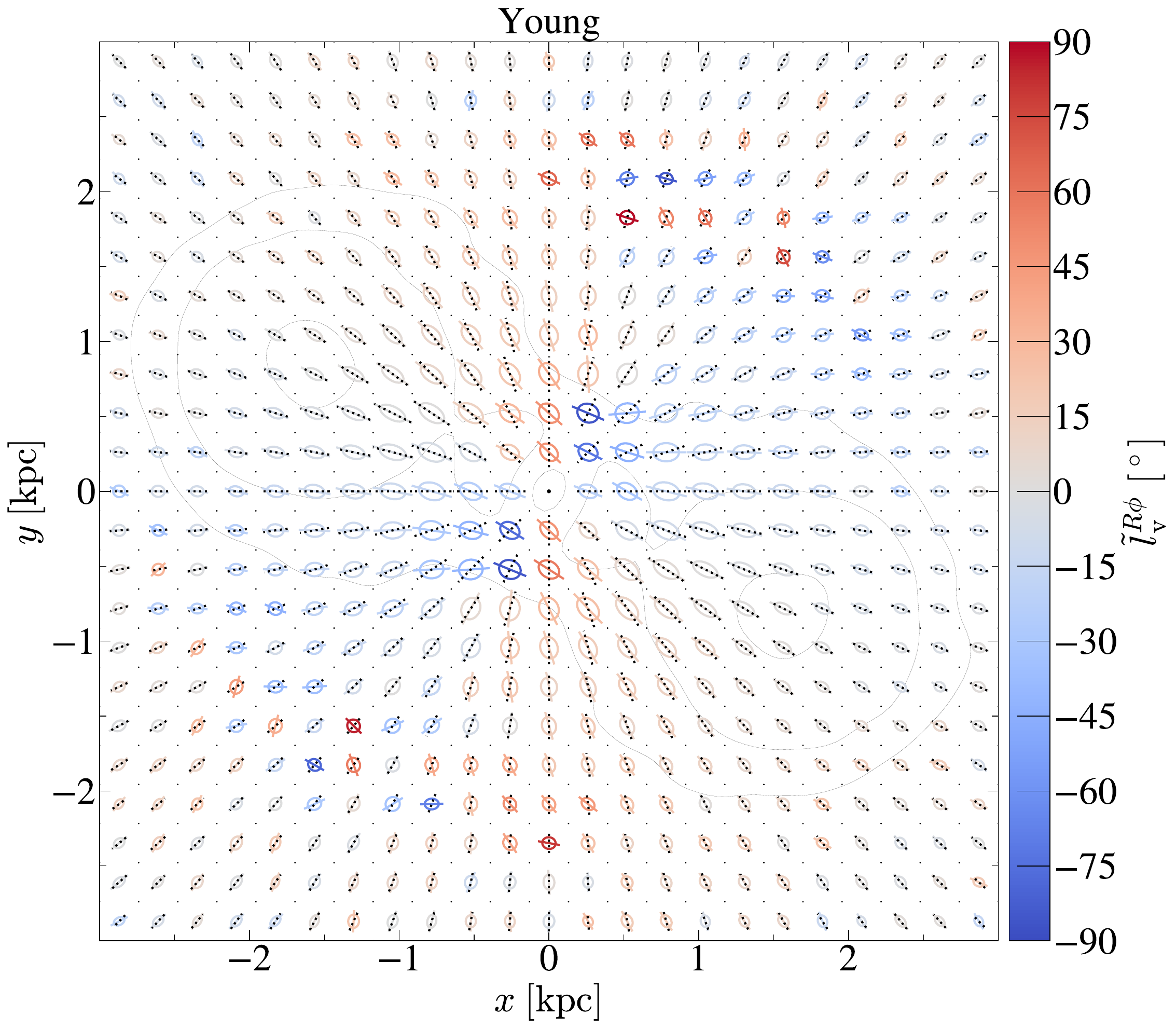}
    \caption{Velocity ellipses of the young population, as previously shown in the left panel of Fig.~\ref{fig:xy_ellipses}, this time colour-coded by the vertex deviation computed using Eqn.~\ref{eq:atan2}, whose values live in the range $(-90,90]\degrees$.}
    \label{fig:vertex_noabs_ellipses}
\end{figure}

To illustrate the comparison between the two vertex deviation definitions, in Fig.~\ref{fig:vertex_noabs_ellipses} we show the same velocity ellipses as in Fig.~\ref{fig:xy_ellipses}, this time colour-coded by the values resulting from Eqn.~\ref{eq:atan2}. The inner regions along the bar's semi-minor axis, where $\ani<0$, are clearly highlighted, reaching values close to $-90\degrees$ where the velocity ellipses are perpendicular to $\Rhat$.

Eqn.~\ref{eq:standard_abs} should be used when we do not want vertex deviation to encode any information about which velocity dispersion dominates. For example, $\vertex$ values of 0 and $90\degrees$ returned by Eqn.~\ref{eq:atan2} both describe a velocity ellipse which is not tilted with respect to the coordinate axes, which means the covariance is negligible relative to the magnitude of the dispersion difference. Eqn.~\ref{eq:standard_abs} would treat both of these cases the same by assigning them a value of $\vertexabs=0\degrees$. The information about which dispersion dominates, \ie\ which of the two coordinate axes the ellipses' semi-major axis is closest to, is quantified by the anisotropy.

\subsection{Caution on error treatment}

It is worth noting that, when estimating uncertainties using bootstrapping, Eqn.~\ref{eq:bootstrap} must be applied with some care when using Eqn.~\ref{eq:atan2} to compute vertex deviation. This is because of the need to limit the range of $\mathrm{atan2}$ to a particular branch so as to make the values unique. The smallest of the two angles which result from the crossing of two straight lines cannot be greater than $90\degrees$. Using the bootstrap method to compute $\upepsilon(\vertex)$ can produce bootstrap values which differ from the value computed from the original population by more than $90\degrees$, given the principal branch spans a region of $180\degrees$. This would lead to an overestimation of the error. The solution is to take those bootstrap ellipses and consider the other end of their semi-major axis, reflecting their $\vertex^*$ values across the origin. Calling the vertex deviation of the original population $\vertex$, the resulting bootstrap distribution belongs to the branch ${(\vertex - 90, \vertex + 90]\degrees}$, where for any bootstrap value $\vertex^*$ we would have ${|\vertex^*-\vertex|\leq 90\degrees}$, as it should be.

\subsection{Relation to the anisotropy and correlation}
\label{appendix:anicorr}

Eqn.~\ref{eq:standard} can be expressed in terms of the correlation ($\corrij$) and in-plane anisotropy ($\aniij$), defined in Eqn.~\ref{eq:corr} and \ref{eq:anisotropy} respectively, as
\begin{equation}
    \tan(2\vertexabs) = 2\corrij \frac{\sqrt{1-\aniij}}{\aniij},
    \label{eq:standard_anicorr}
\end{equation}
so that the equivalent of Eqn.~\ref{eq:standard_abs} is 
\begin{equation}
    \vertexabs = \frac{1}{2}\arctan\frac{2\corrij \sqrt{1-\aniij}}{\left| \aniij \right|}.
\label{eq:standard_abs_anicorr}
\end{equation}
The vertex deviation using the equivalent of Eqn.~\ref{eq:atan2} instead is
\begin{equation}
    \vertex = \frac{1}{2}\mathrm{atan2}\left(2\corrij \sqrt{1-\aniij},\aniij\right).
    \label{eq:atan2_anicorr}
\end{equation}

\section{Additional models}
\label{appendix:weaker_bars}

\begin{figure}
    \centering
    \includegraphics[width=\columnwidth]{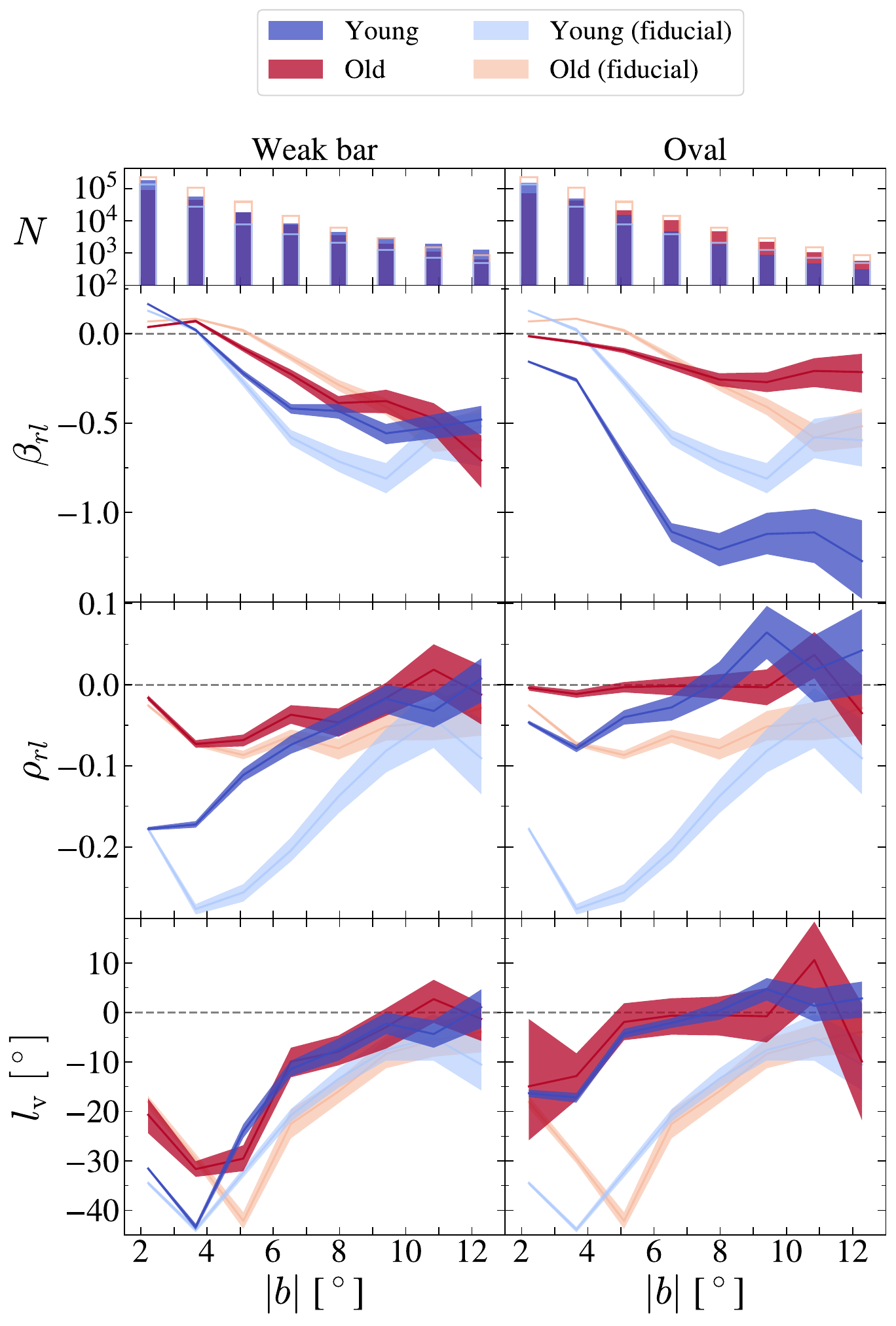}
    \caption{Anisotropy (top), correlation (middle) and vertex deviation (bottom) as a function of latitude along the minor axis of the bulge, $|l|<2\degrees$, within ${\Rgc<3.5\kpc}$, for models with a weaker bar (left panels) and only a central oval (right panels), compared with the strongly-barred fiducial model (light colours). We split the stellar populations, and scale the new models, as described for the fiducial model in the main text. The surfaces represent the 68\% percentile bootstrap confidence intervals.}
    \label{fig:othermodels}
\end{figure}
In Fig.~\ref{fig:othermodels} we compare the anisotropy, correlation, and vertex deviation of our fiducial model with two additional models. The first model is the weak bar model presented by \citet{gough-kelly2022}. The second model has not previously been presented; it has only a central weak oval, not a bar. We scale the three models identically, with the bar/oval orientated as in the fiducial model, and using the same age cuts for the young and old stellar populations. 

We select stars along the bulge minor axis, namely ${|l|<2\degrees}$ and ${\Rgc<3.5\kpc}$. The figure shows that the amplitude of \ani\ is largest in the young population of the central-oval model, not in the fiducial model, which in turn has a larger amplitude than in the weak bar's young population. In the vertex deviation, the (negative) amplitude of the young population is comparable between the fiducial and weak bars, while in the weak oval case the amplitude of \vertexabs\ in smaller than in the other two models for both populations. 

In contrast with \ani\ and \vertexabs, the correlation has amplitudes that follow the strength of the bar in the ${3\degrees<|b|<6\degrees}$ region, with the largest \corr\ in the young population of the fiducial model, and the smallest (near-zero) \corr\ in the old population of the weak oval model. From these results we conclude that the correlation is the best tracer of the bar strength of any given population.

\section{Testing the assumption of bootstrapping}
\label{appendix:bootstrap_assumption}

In this subsection we use the model to briefly discuss the validity of bootstrapping for the kinematic variables of interest.

Let us suppose we have a sample of size $n$, extracted from a larger population of ${\mathrm{size} \gg n}$. We are interested in knowing how closely the correlation of the sample, called the sample estimate $\hat{\rho}$, approximates the correlation of the total population, $\rho$. If we had access to the population, we could investigate this by drawing many random samples of size $n$ with replacement from the population, and computing all their sample estimates $\hat{\rho}$. The histogram resulting from binning these values is called the sampling distribution. Its standard deviation, $\sigma(\hat{\rho})$, is called the standard error, and measures the precision of the sample estimates, quantifying their variability around the mean, ${\langle \hat{\rho} \rangle}$. The difference between the mean and the value computed from the population, ${\langle \hat{\rho} \rangle - \rho}$, is called the bias, and it measures the average accuracy of the sample estimates. If the bias is zero, the standard error measures both the precision and accuracy of the sample estimates, as it quantifies their variability around the true value, computed from the population.

\begin{figure*}
    \begin{subfigure}{0.45\textwidth}
        \centering
        \includegraphics[width=0.97\textwidth]{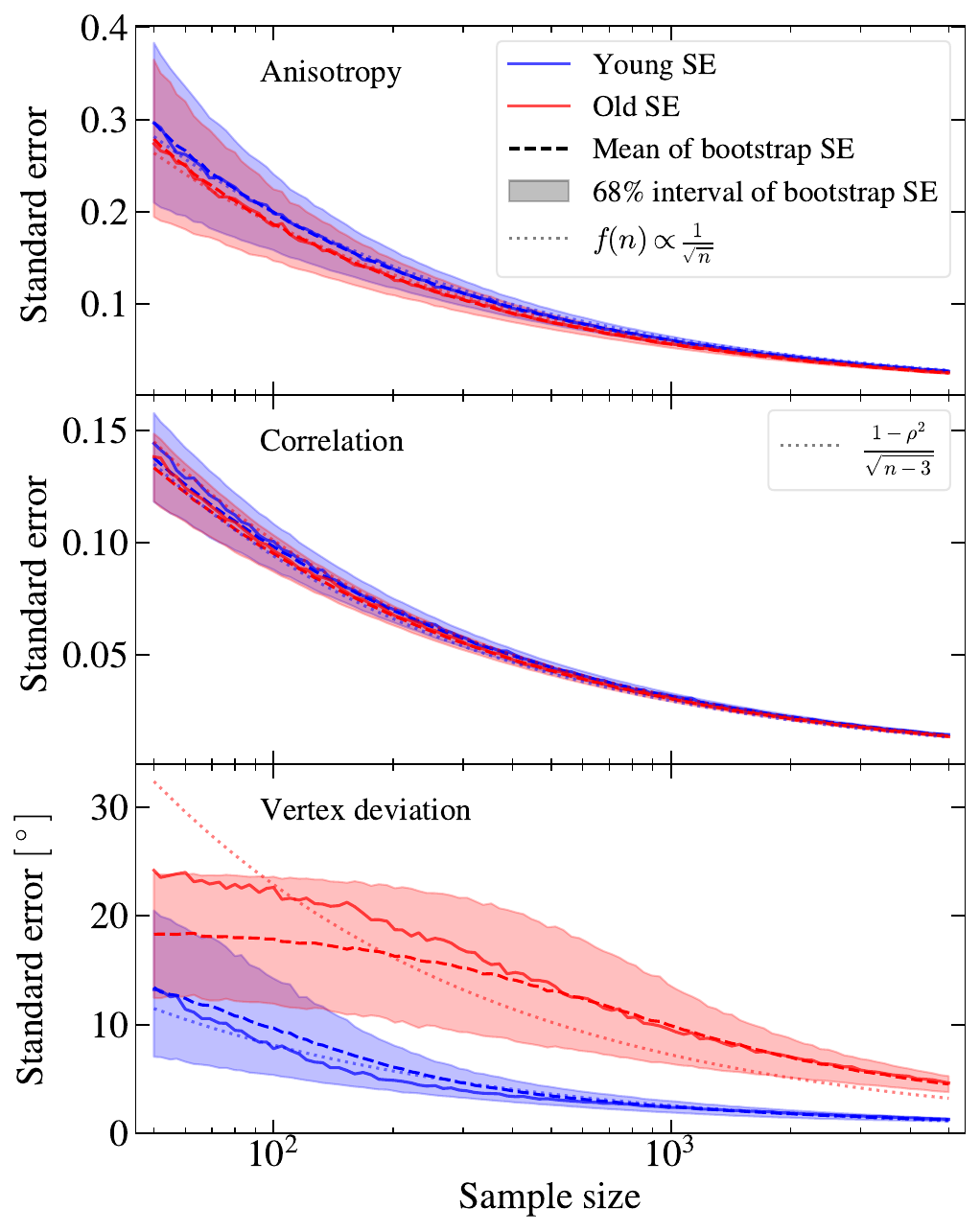}
        \caption{Standard error}
        \label{subfig:standard_error}
    \end{subfigure}
    \begin{subfigure}{0.45\textwidth}
        \centering
        \includegraphics[width=\textwidth]{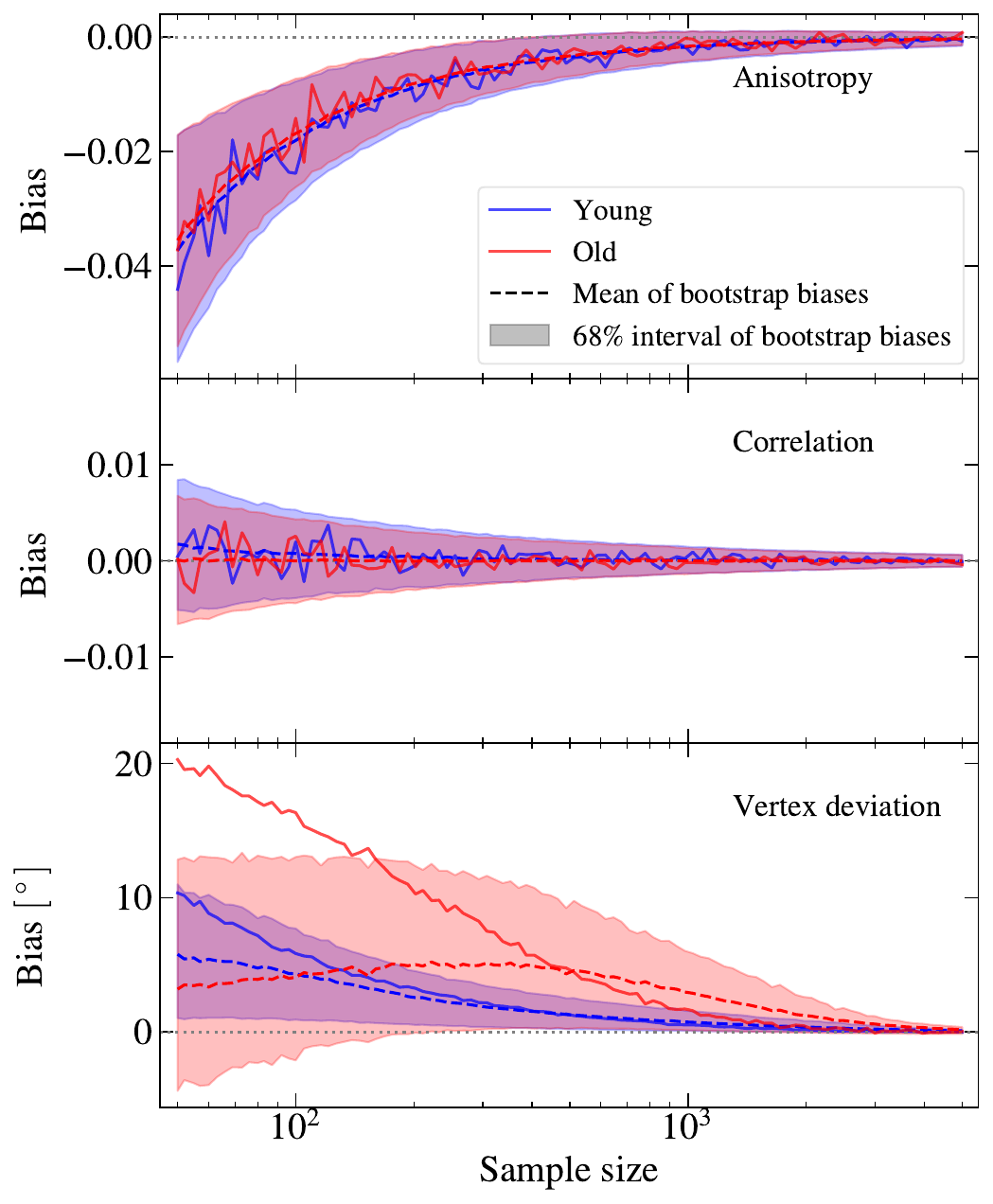}
        \caption{Bias}
        \label{subfig:bias}
    \end{subfigure}
    \caption{Standard error (a) and bias (b) as a function of sample size, for young (blue) and old (red) stars on the bulge minor axis. The dashed lines show the mean bootstrap standard error (a) and mean bootstrap bias (b). The variability of bootstrap standard errors and bootstrap biases around the mean is illustrated by the $68\%$ percentile range surface. On column (a) the dotted curves show fits of $f(n) \propto n^{-1/2}$ to the standard errors. On column (b) the horizontal dotted lines indicate zero.}
    \label{fig:bootstrap_test}
\end{figure*}

In inferential statistics, we do not have access to the population, only to the sample, of size $n$. Therefore, we cannot compute the precision or accuracy of the sample estimate, $\hat{\rho}$, as described above because we cannot build the sampling distribution. Instead, we can take many \textit{bootstrap} samples of size $n$ with replacement from the sample itself, and compute all their correlation values, $\rho^*$, called bootstrap estimates. The histogram resulting from binning these values is called the bootstrap distribution. Its standard deviation, $\sigma(\rho^*)$, is called the bootstrap standard error, and the difference between its mean and the original sample estimate, ${\langle \rho^* \rangle - \hat{\rho}}$, is called the bootstrap bias. If the bootstrap assumption held true, meaning the original sample was representative of the underlying population, then the bootstrap standard error would approximate the actual standard error, \ie\ ${\sigma(\rho^*)\approx \sigma(\hat{\rho})}$, and the bootstrap bias would approximate the actual bias, \ie\ ${\langle \rho^* \rangle -\hat{\rho} \approx \langle \hat{\rho} \rangle -\rho}$. Therefore, we would be able to use these quantities as a measure of the precision and accuracy of our original sample estimate \citep{hesterberg2015}.

We use our model to test the bootstrap assumption on our statistics of interest at different sample sizes for the young and old populations. We select the stars on the bulge minor axis, namely ${|l|<2^\circ}$, ${3^\circ<|b|<6^\circ}$ and ${\Rgc<3.5\kpc}$. From each population we extract $5000$ random samples with replacement of sample sizes varying from $n=50$ to $5000$. Samples of size $n=5000$ represent $14.8\%$ and $3.5\%$ of the young and old populations respectively. For each sample size we aggregate the $5000$ sample estimates of the statistics of interest ($\hat{\ani}$, $\hat{\corr}$ and $\hat{\vertexabs}$) into a sampling distribution, and we compute the standard error and bias as described in the first paragraph. Therefore, we obtain a value of standard error and bias for each sample size $n$, shown as the solid lines in Fig.~\ref{fig:bootstrap_test}.

From each of the $5000$ samples of size $n$ extracted from a population, we then draw, from the sample itself, $500$ bootstrap samples of the same size $n$ with replacement, and aggregate their values of the statistics of interest, \ie\ the bootstrap estimates ($\ani^*$, $\corr^*$ and $\vertexabs^*$), into a bootstrap distribution. We then compute the bootstrap standard error and the bootstrap bias as described above. Doing this for all $5000$ samples of size $n$ extracted from the population results in $5000$ values of bootstrap error and bootstrap bias. We take the average and show the mean bootstrap standard error and mean bootstrap bias as dashed lines in Fig.~\ref{fig:bootstrap_test}. The $68\%$ percentile range is also shown as a shaded region.

In Fig.~\ref{subfig:standard_error} we show the standard error as a function of sample size. For the anisotropy and correlation in the first two panels, the standard error (solid line) and mean of the bootstrap standard errors (dashed line) match closely for both populations for all $n$. Therefore, on average the bootstrap assumption holds for these variables. Given the standard error curves for the young and old populations match quite closely despite their true values of anisotropy and correlation differing (see Fig.~\ref{fig:kinematics_lat}), these conclusions depend almost solely on sample size. We illustrate this with the dotted lines, which show fits of $f(n) \propto n^{-1/2}$ to the actual standard error curves. For the correlation we show the analytic expression recommended by \citet{Gnambs_2023}, which depends on the correlation value itself, though this makes little difference in practice at our range of values ($\lesssim 0.3$).

The standard error and mean bootstrap standard error of the vertex deviation of the young stars also roughly match, with maximum deviation of ${\sim}2\degrees$. However, for the old stars at sample sizes below $n\lesssim 400$, these two curves separate, with the mean bootstrap standard error underestimating the actual standard error by up to ${\sim}6\degrees$. The $68\%$ interval of the old population also takes longer to converge to the actual standard error. Moreover, unlike in the other panels, the standard errors for the young and old populations are different for all sample sizes, and hence do not depend only on $n$. The vertex deviation of isotropic and weakly correlated populations, like the old one here (see Fig.~\ref{fig:kinematics_lat}), whose velocity ellipse is closer to a circle with less well-defined semi-major axis direction, have larger standard errors and the bootstrap assumption on average breaks below a certain $n$, with the bootstrap error on average underestimating the actual standard error.

In Fig.~\ref{subfig:bias} we show the bias as a function of sample size. For anisotropy and correlation, the bias (solid line) and the mean bootstrap bias (dashed line) match closely for both populations for all $n$, which again confirms that the bootstrap assumption holds for these statistics. It is worth noting that the anisotropy develops a negative bias that increases in magnitude as sample sizes drop below $n\approx 10^3$. This means that the sample estimates at those sample sizes on average deliver an anisotropy value lower than the true value computed from the population. However, the effect is small ($\lesssim0.04$) even for the smallest samples of $n=50$. Therefore, the correlation and anisotropy are largely unbiased estimators, which means that the standard error of these statistics is a measure of both the precision and accuracy of sample estimates, as it measures their variability around the true value, computed from the population.

The bootstrap assumption again breaks for vertex deviation, this time for both young stars below $n\approx 200$ and old stars below $n\approx 400$, reaching a discrepancy of ${\sim}5\degrees$ and ${\sim}17\degrees$ respectively between the actual bias and the mean bootstrap bias at the smallest sample size of $50$. For both populations the mean bootstrap bias underestimates the actual bias. Moreover, unlike the anisotropy and correlation, the vertex deviation develops a significant bias below $n\lesssim 10^3$, reaching up to $10^\circ$ and $20^\circ$ for the young and old populations respectively at $n=50$. As a result, the standard error of vertex deviation at these sample sizes is a measure of the precision of the sample estimates but not of their accuracy in representing the true value of the underlying population.

We summarise the findings of this appendix below:
\begin{enumerate}
    \item The standard error and bias of the anisotropy and correlation depend almost solely on sample size, $n$. The standard error varies as $\propto n^{-1/2}$, and only small anisotropy biases are introduced at low $n$. For vertex deviation, the standard error and bias at the same $n$ differ for the young and old populations, with larger standard errors and biases for the old. Vertex deviation develops biases for both the young and old populations at $n\lesssim 10^3$, reaching ${\sim}5\degrees$-$15\degrees$ at $n=100$. This means that at those sample sizes the standard error quantifies the precision of vertex deviation estimates but contains limited information about their accuracy in representing the underlying population.
    \item The bootstrap assumption holds on average for the anisotropy and correlation at all $n$, with the mean bootstrap standard error and bias matching the actual standard error and bias respectively. The same is roughly true for the standard error of the vertex deviation of the young stars, but not for the old, whose mean bootstrap standard error underestimates the actual standard error below $n\approx400$ by up to ${\sim} 6 \degrees$. Below that $n$, bootstrapping also underestimates the bias by up to $17\degrees$ for the old stars and $5\degrees$ for the young at ${n=50}$.
\end{enumerate}
%
%

\bsp	
\label{lastpage}
\end{document}